# An integrated approach to soil structure, shrinkage, and cracking in samples and layers

V.Y. Chertkov*

Division of Environmental, Water, and Agricultural Engineering, Faculty of Civil and Environmental Engineering, Technion, Haifa 32000, Israel

**Abstract.** A recent model showed how a clay shrinkage curve is step-by-step transformed into the shrinkage curve of an aggregated soil at any clay content if it is measured on samples so small that cracks do not occur at shrinkage. Such a shrinkage curve was called a reference curve. The present work generalizes this model to any soil sample size or layer thickness, i.e., to any crack contribution to the shrinkage curve. The approach is based on: (i) recently suggested features of an intra-aggregate structure; (ii) detailed accounting for the contributions to the soil volume and water content during shrinkage; and (iii) new concepts of lacunar factor, crack factor, and critical sample size. The following input parameters are needed for the prediction: (i) all parameters determining the basic dependence of the reference shrinkage curve; (ii) parameters determining the critical sample size (structural porosity and minimum and maximum aggregate size at maximum swelling); and (iii) initial sample size or layer thickness. A primary experimental validation of the new model concepts is conducted using the relevant available data on the shrinkage curves of four soils with different texture and structure that were obtained utilizing the samples of two essentially different sizes. The results show evidence in favor of the model.

*Corresponding author. Tel.: 972-4829-2601.
E-mail address: agvictor@tx.technion.ac.il; vychert@ymail.com (V.Y. Chertkov).

## 1. Introduction

Physical understanding of soil shrinkage and cracking is critically important for similar understanding of the hydraulic properties, water flow, and transport phenomena in soils. The major issue can be formulated as follows: how can one express the crack volume at soil shrinkage and the shrinkage curve of the soil with cracks through inter- and intra-aggregate soil structure, clay content and type, as well as sample size or layer thickness (here and below we imply the physical prediction without fitting). At present time the total physical understanding of the issue is lacking, although there are various data (which may be insufficient) and some theoretical results (see References below). There are experimental estimates of soil crack volume (e.g., Zein el Abedine and Robinson, 1971; Yaalon and Kalmar, 1984; Dasog et al., 1988). There are also models connecting the crack volume with shrinkage curve (e.g., Chertkov and Ravina, 1998, 1999; Chertkov, 2000a). Nonetheless, the shrinkage curve is found from experimental data, but not through parameters of soil structure, clay content and type. Except for that the sample size effects were not considered in the modeling. Thus, it is currently impossible to physically predict the crack volume. There are many works devoted to the measurement of the soil shrinkage curve using different methods, sample sizes, and soils with the relatively small organic matter content, 0.15-3.5% by weight (e.g., McGarry and Daniels, 1987; Tariq and Durnford, 1993; Olsen and Haugen, 1998;

Crescimanno and Provenzano, 1999; Braudeau et al., 2004, 2005; Peng and Horn, 2005; Boivin et al., 2006; Cornelis et al., 2006). These works constate that in general case the shrinkage curve of a cracked soil depends on soil structure and sample size, and propose the different fitting approximations of the shrinkage curves. At sufficiently small sample size the shrinkage curve does not depend on the latter and contains no cracks (more accurately their volume is negligible) (e.g., Crescimanno and Provenzano, 1999; Braudeau et al., 2004; Boivin et al., 2006). However, where does the border lie between "small" and "large" samples from the physical viewpoint? There is also the physical model of the so-called reference shrinkage curve, i.e., the shrinkage curve that is obtained using the sufficiently small samples when the crack volume is negligible (Chertkov, 2007a, 2007b, 2007c, 2008a). This model links the reference shrinkage curve with soil structure and physical properties and quantitatively explains the shape of the shrinkage curve. However, the possibility to physically predict the shrinkage curve while accounting for the sample size, and correspondingly, cracking is so far lacking. In addition, the soil shrinkage curve and crack volumes for geometry of layer (i.e., for the actual field conditions) and sample are quite different (Chertkov, 2005a).

Below we propose the solution of the above issue. Detailing the title of this work, its *objective* is to consider without fitting the *effects* of clay content, clay type, inter- and intra-aggregate soil structure, sample size and layer thickness, on such *characteristics of soil shrinkage* as evolution of crack, matrix, and soil volume with water content decrease in the framework of an *integrated approach*. The *first* major point of the approach to be used is the recently suggested intra-aggregate structure accounting for lacunar pores and an aggregate surface layer with specific properties (Chertkov, 2007a, 2007c, 2008a). The *second* major point is the consideration and discussion of all the contributions to the soil volume and water content based on the inter- and intra-aggregate soil structure. The *third* major point is the use of new concepts of *lacunar factor* ($k$), *crack factor* ($q$), and *critical sample size* ($h^*$). The content of the theory is reflected by the titles of Sections 2-8. We consider in detail the physical soil parameters that are needed to check the approach (Section 9). Finally, we analyze the relevant available data in order to check the different aspects of the approach (Sections 10 and 11). Notation is summarized at the end of the paper.

**2. Water content and volume balance equations based on the soil structure**

The objective of this section is to derive simple equations that determine the variation of a soil volume as water content decreases, accounting for the actual contributions to the soil volume and water content. We rely on the structure of aggregated soils with negligible organic matter content from Chertkov (2007a, 2007c, 2008a). Figure 1 illustrates the soil structure in detail. The key point of the structure is the existence of the surficial (or interface) aggregate layer that is deformable, but does not shrink. The introduction of this layer allows one to quantitatively (without fitting) explain the observed reference shrinkage curve (Chertkov 2007a, 2007c, 2008a) and the soil-water retention at drying (Chertkov, 2010a).

The gravimetric water content of the soil, $W$ includes two contributions (Chertkov 2007a, 2007c, 2008a): the water content of the intra-aggregate matrix, $w'$ (per unit mass of the oven-dried soil) and the water content of the interface layer, $\omega(w')$ (Fig.1)

$$W(w')=w'+\omega(w'), \quad 0 \leq w' \leq w'_h, \quad 0 \leq W \leq W_h \tag{1}$$

This water content balance equation suggests that inter-aggregate pores are large compared to intra-aggregate pores. For this reason at the maximum swelling point, $W_h$ (Fig.2) when aggregates are water saturated the inter-aggregate pores from the capillarity considerations stay empty and do not contribute to the soil water content, $W$ (Eq.(1)) at $W<W_h$, i.e., at shrinkage. It is obvious (Fig.1) that the specific volume of the soil with cracks, $Y$ consists of two contributions, the specific volume of aggregates (per unit mass of the oven-dried soil), $U_a(w')$ and specific volume of cracks (per unit mass of the oven-dried soil), $U_{cr}(w')$

$$Y(w')=U_a(w')+U_{cr}(w'), \quad 0 \leq w' \leq w'_h \qquad (2)$$

The growth of the $U_{cr}(w')$ volume is the result of the evolution of inter-aggregate pores at shrinkage, starting from the initial value, $U_s=U_{cr}(w'_h)$ that corresponds to these pores at maximum swelling point $w'=w'_h$. In turn, the $U_a(w')$ volume also includes two contributions (Fig.1): the specific volume of the interface layer (per unit mass of the oven-dried soil), $U_i$ that does not depend on $w'$, and the specific volume of the intra-aggregate matrix (per unit mass of the oven-dried soil), $U'(w')$

$$U_a(w')=U_i+U'(w'), \quad 0 \leq w' \leq w'_h \qquad (3)$$

Joining Eqs.(2) and (3) one obtains the *first* volume balance equation of the soil at shrinkage

$$Y(w')=U_i+U'(w')+U_{cr}(w'), \quad 0 \leq w' \leq w'_h \qquad (4)$$

This equation relates to the soil as a whole. The *second* volume balance equation relates to the intra-aggregate matrix and presents $U'(w')$ as the sum of the specific volumes (per unit mass of the oven-dried soil) of soil solids (clay particles, silt, and sand grains), $U'_{cs}$ that does not depend on $w'$, clay pores, $U'_{cp}(w')$, and lacunar pores, $U'_{lp}(w')$ (Fig.1)

$$U'(w')=U'_{cs}+U'_{cp}(w')+U'_{lp}(w'), \quad 0 \leq w' \leq w'_h \ . \qquad (5)$$

The specific volume, $U(w)$ of the intra-aggregate matrix per unit mass of the oven-dried matrix itself ($w$ is the water content of the intra-aggregate matrix per unit mass of the oven-dried matrix itself) is connected with $U'(w')$ as (Chertkov 2007a) (Fig.2)

$$U'(w')=U(w)/K, \quad w'=w/K \qquad (6)$$

where $K>1$ is the aggregate/intra-aggregate mass ratio. Similar relations can be written for the values of $U_{cs}$ ($\equiv 1/\rho_s$; $\rho_s$ being the density of solids), $U_{cp}(w)$, and $U_{lp}(w)$ (related to unit mass of the oven-dried intra-aggregate matrix itself)

$$U'_{cs}=U_{cs}/K, \quad U'_{cp}(w')=U_{cp}(w)/K, \quad U'_{lp}(w')=U_{lp}(w)/K \ . \qquad (7)$$

Equation (5) in terms of $U$, $U_{cs}$, $U_{cp}(w)$, and $U_{lp}(w)$ (instead of the values with a prime) has been used (Chertkov, 2007a, 2007c) in considering of the *reference* shrinkage curve at any clay content when the specific volume of the inter-aggregate (structural) pores, $U_s$ (Fig.1) is kept at shrinkage, and cracks do not appear, i.e., when

$$U_{cr}(w')=U_s=\text{const} \ . \qquad (8)$$

In this case as it can be seen from Eqs.(4) and (8) that the soil shrinkage curve $Y(w')$ as a function of $w'$ (or $W$ according to Eq.(1)) is only determined by the $U'(w')$ dependence, and the second volume balance equation (Eq.(5)) plays the major part. Unlike that in this work, we consider the general case of an arbitrary sample size (and layer thickness) and the corresponding $U_{cr}(w')$ dependence. Here, the first volume balance equation (Eq.(4)) and its role in the task solution become complicated.

**3. Definition of lacunar ($k$) and crack ($q$) factors**

The objective of this section is to introduce two factors that characterize the intra- and inter-aggregate variations of the soil volume at drying and allow one (see section 4) to find dependencies $Y(w')$ and $U_{cr}(w')$ based on Eqs.(4) and (5). Equation (5) corresponds to shrinkage of the intra-aggregate matrix (Fig.1) while Eq.(4) corresponds to shrinkage of the soil as a whole (Fig.1). However, both these equations have an *analogical* structure. The first term in the right part ($U_i$ in Eq.(4) and $U''_{cs}$ in Eq.(5)) is a constant as water content decreases. The second term is a volume whose variations initiate both the variations of the third term and the term in the left part. For instance, the primary variations of $U''_{cp}(w')$ in Eq.(5) lead to secondary variations of $U''_{lp}(w')$ and $U'(w')$. Similarly, the primary variations of $U'(w')$ in Eq.(4) lead to the secondary variations of $U_{cr}(w')$ and $Y(w')$. This connection between the primary and secondary volume variations becomes more clear if Eqs.(4) and (5) are written in differential form as

$$dY(w')=dU'(w')+dU_{cr}(w'), \quad 0\leq w'\leq w'_h , \tag{4'}$$

$$dU'(w')=dU''_{cp}(w')+dU''_{lp}(w'), \quad 0\leq w'\leq w'_h . \tag{5'}$$

Equation (5') has been considered in the case of crack absence (Chertkov, 2007c, 2008a) for the values without a prime (see Eqs.(6) and (7)). In these works the link between the primary volume variation ($dU''_{cp}(w')$) and secondary ones ($dU''_{lp}(w')$ and $dU'(w')$) was determined by lacunar factor, $k$. *By definition*, $k$ is the fraction of the increment of the clay matrix pore volume at shrinkage, $dU''_{cp}(w')<0$ that is transformed to the corresponding increment of the lacunar pore volume inside the intra-aggregate matrix, $dU''_{lp}(w')>0$ (Fig.1). That is, by definition

$$dU''_{lp}(w')=-k\, dU''_{cp}(w'), \quad 0\leq k\leq 1, \quad 0\leq w'\leq w'_h . \tag{9}$$

Then, from Eq.(5') the increment of the intra-aggregate matrix volume (and increment of the aggregate volume because $U''_{cs}$=const in Eq.(5) and $U_i$=const in Eq.(4)) is

$$dU'(w')=(1-k)\, dU''_{cp}(w'), \quad 0\leq k\leq 1, \quad 0\leq w'\leq w'_h . \tag{10}$$

$k$ is the characteristics of the soil that depends on clay type and clay content, $c$, but not on water content (Chertkov, 2007c, 2008a). At a high clay content, $c>c_*$ ($c_*$ is the critical clay content (Chertkov, 2007a)) the lacunar pore volume (Fig.1) is negligible and according to Eq.(9) $k=0$ in the whole range $0\leq w'\leq w'_h$. At $c<c_*$ lacunar pores exist, and according to Eqs.(9) and (10) $0<k<1$. The following result (Chertkov, 2007c, 2008a) is essential: $k$, that by definition, is connected with the variation of the intra-aggregate structure of a soil at shrinkage, simultaneously determines the slope $S_r$ of the reference shrinkage curve in the basic shrinkage range as (Fig.2)

$$S_r\equiv dU'/dW=(1-k)/\rho_w, \quad W_n\leq W\leq W_s \tag{11}$$

where $\rho_w$ is the water density, and $W_n$ and $W_s$ are the end-points of the basic and structural shrinkage, respectively. For pure clays and soils with high clay content $k=0$ and the slope is numerically equal to unity (Chertkov, 2007a). Equation (11) shows the simple link between the immediately observed (on small samples; for exact meaning of the terms "small" and "large" see section 6) *macro-parameter* of reference soil shrinkage ($S_r$) and *micro-parameter* of the intra-aggregate structure ($k$). Since $k$ is connected with the intra-aggregate structure and does not depend on inter-aggregate pores (that are transformed to cracks), the expression $k(c)$ (section 5) and definition of $k$ (Eqs.(9) and (10)) are kept in the general *case of crack development* in sufficiently large samples. However, compared to Eq.(11), the expression for the slope, $S$ of the shrinkage curve in the basic shrinkage range in this case changes (Section 4).

Now, in the force of analogy between Eqs.(4') and (5') (see beginning of this section), one can introduce a crack factor $q$ and, based on Eq.(4'), express the secondary volume variations, $dY(w')$ and $dU_{cr}(w')$ through the primary one, $dU'(w')$ as

$$dU_{cr}(w') = -q\, dU'(w'), \quad 0 \leq q \leq 1, \quad 0 \leq w' \leq w'_h, \tag{12}$$

$$dY(w') = (1-q)\, dU'(w'), \quad 0 \leq q \leq 1, \quad 0 \leq w' \leq w'_h. \tag{13}$$

That is, *by definition* $q$ is the fraction of the increment of the aggregate volume at shrinkage, $dU_a(w')=dU'(w')<0$ (see Eq.(3) where $U_i$=const) that is transformed to the corresponding increment of the crack volume inside the soil, $dU_{cr}(w')>0$. We *assume* that $q$ does not depend on water content (cf. with similar assumption for $k$ in Chertkov, 2007c, 2008a). This *assumption* 1 will be justified, by the available data (section 11.3). The dependence of $q$ on the sample size is considered in Section 7.

**4. Specific soil and crack volumes as functions of lacunar and crack factors**

Integrating Eq.(12) with the initial condition $U_{cr}(w'_h)=U_s$ and replacing $U'(w')$ with $U(w)/K$ (Eq.(6)), one can express the specific crack volume, $U_{cr}(w')$ through the specific volume of the intra-aggregate matrix, $U(w)$ as (see $U_h$ and $U(w)$ in Fig.2 and $w'(w)$ in Eq.(6))

$$U_{cr}(w') = q\,(U_h - U(w))/K + U_s \tag{14}$$

Replacing in Eq.(4) $U_{cr}(w')$ from Eq.(14) and $U'(w')$ from Eq.(6) one can also express the specific soil volume $Y(w')$ through $U(w)$ as

$$Y(w') = (1-q)U(w)/K + qU_h/K + U_s + U_i. \tag{15}$$

Thus, the soil (Eq.(15)) and crack (Eq.(14)) volumes as functions of water content are determined by the intra-aggregate matrix volume, $U(w)$ and the water content relations (Eqs.(1) and (6)). Note that in the case of the sufficiently small samples, when the inter-aggregate pores do not grow and are not transformed to cracks (Eq.(8); Fig.1), i.e., $q=0$ (see Eq.(12)), the specific soil volume, $Y(w')$ from Eq.(15) is reduced to the reference shrinkage curve, $Y(w') \equiv Y_r(w') = U(w)/K + U_s + U_i$ (together with Eq.(1)) that was considered in Chertkov (2007a, 2007c, 2008a). Finding the $U(w)$ dependence for the specific volume of the intra-aggregate matrix (as well as the constant specific volumes, $U_h$, $U_i$, and $U_s$) has been considered in detail in these works. In particular, the essential dependence of $U(w)$ on the lacunar factor, $k$ has also been considered. This dependence of $U(w)$ on $k$ leads to corresponding dependences of $U_{cr}(w')$ (Eq.(14)) and $Y(w')$ (Eq.(15)). Hence, as in the case of the reference shrinkage curve, in the general case $Y(w')$ in Eq.(15) and $U_{cr}(w')$ in Eq.(14) together with Eq.(1) give the parametric presentation of the shrinkage curve, $Y(W)$, as well as the specific crack volume vs. the soil water content, $U_{cr}(W)$ (Fig.2). These presentations allow one to calculate $Y(W)$ and $U_{cr}(W)$ through $U(w)$ or $Y_r(w')$ (accounting for the $U(w)$ dependence on $k$ from Chertkov, 2007a, 2007c, 2008a) and the crack factor, $q$ (for a number of physical soil parameters necessary for the calculation see section 9). The slope $S \equiv dY/dW$ of the shrinkage curve $Y(W)$ of the soil with cracks ($q>0$) in the basic shrinkage range $W_n \leq W \leq W_s$ (see Fig.2) can be found as (Eq.(13))

$$S \equiv dY/dW = (1-q)dU'/dW, \qquad 0 \leq W \leq W_h . \qquad (16)$$

However, in the basic shrinkage range (see Eq.(11)) $dU'/dW=(1-k)/\rho_w$ (Chertkov, 2007c). Hence, for the soil with cracks

$$S=(1-q)(1-k)/\rho_w, \qquad W_n \leq W \leq W_s . \qquad (17)$$

Thus the task of finding the $Y(W)$ and $U_{cr}(W)$ dependences (Eqs.(15), (14), (1), and (6)) and, in particular, the $S$ slope (Eq.(17); Fig.2), is reduced to finding $k$ and $q$.

**5. Lacunar factor expression**

Derivation of the lacunar factor, $k$ was recently considered (Chertkov, 2010b). The objective of this section is to briefly review the derivation results as applied to our aims in this work. The lacunar pores develop inside aggregates (Fig.1) (Fiès and Bruand, 1998). For this reason and according to its physical meaning, the $k$ value can only depend (except for the soil clay content, $c$) on the characteristics of the intra-aggregate matrix (see Fig.1). These characteristics include ones of the contributive clay (reflecting the clay type), the relative volumes of clay solids, $v_s$ and the oven-dried clay, $v_z$ (Chertkov, 2000b, 2003) as well as the porosity $p$ of the contributive silt and sand grains when they are in the state of *imagined* contact. $v_s$, $v_z$, and $p$ enter the expression for the critical clay content, $c_*$ (Chertkov, 2007a) as

$$c_* = [1+(v_z/v_s)(1/p-1)]^{-1} . \qquad (18)$$

In addition, $k$ is a dimensionless value. For this reason it is assumed that at $c<c_*$ the soil lacunar factor $k$ as a function of the clay content, $c$ is a universal function $k(c/c_*)$ of the $c/c_*$ ratio at $0<c/c_*<1$ (at $c/c_*>1$ the lacunar pores are lacking, and $k=0$). Accounting for the obvious conditions: $k(0)=1$, $k'(0)=k''(0)=0$, $k(1)=0$, $k'(1)=k''(1)=-\infty$ (Fig.3), the following simple expression for $k(c/c_*)$ was found

$$k(c/c_*) = [1-(c/c_*)^3]^{1/3}, \qquad 0<c/c_*<1 . \qquad (19)$$

This expression was validated using available data on sixteen soils from Braudeau and Mohtar (2004), Braudeau et al. (2005), and Boivin et al. (2006).

**6. The critical sample size**

In section 7 we consider the case of a three-dimensional sample with finite sizes. The three sample sizes are usually close. For this reason the samples of approximately cubic shape with side $h$ are considered below. This section is devoted to introducing some sample size, $h^*$ that is critical from the viewpoint of the appearance of cracks in the sample at shrinkage. The preliminary introduction of $h^*$ is necessary for the consideration (in section 7) of the crack factor $q$ as a function of sample size.

Two physical conditions determine the existence and value of such a critical sample size $h^*$. Before formulating these conditions we expose some heuristic considerations in their favor. Any sample consists of aggregates. The cracks in a sample appear as a result of the inter-aggregate pores developing at shrinkage (Fig.1). The aggregates (and the inter-aggregate pores associated with them) are distributed by size. The cumulative aggregate-size distribution at maximum soil swelling, $F(X, X_{min}, X_m, P_h)$ (Chertkov, 2005b) depends on the following parameters: minimum ($X_{min}$) and maximum ($X_m$) aggregate sizes, and inter-aggregate porosity at maximum swelling ($P_h$). Using the $F$ distribution one can estimate the mean distance, $l$ between the aggregates of size $X$, in particular, the mean distances $l_{min}$ and $l_m$ at $X=X_{min}$ and $X=X_m$. Usually $l_{min} \ll l_m$ and $l_m/l_{min} \sim 10^2$. Therefore one can introduce the size $h^*$ that meets the following conditions

$$l_{min} \ll h^* < l_m \ . \tag{20}$$

We are interested here in the samples of such size. Indeed, with high probability samples so small ($h^* < l_m$) do not contain or only contain one large aggregate (large inhomogeneity) and are relatively homogeneous from the viewpoint of the aggregate size range. Therefore, according to fracture mechanics (e.g., Gdoutos, 1993) initiation and development of cracks in such samples under loading (in our case under action of shrinkage stresses) is suppressed compared with that in the larger samples of the same soil. On the other hand, according to Eq.(20) such samples (with size $h^* < l_m$) are simultaneously sufficiently large ($l_{min} \ll h^*$) to still contain the enormous number of aggregates and, hence, to be representative for the soil.

Thus, the physical conditions determining the critical sample size, $h^*$ from the viewpoint of the crack development in the sample at shrinkage are as follows.
(i) Condition of approximate *sample homogeneity*. The number of large aggregates (inhomogeneities) in the sample that lead to cracking at shrinkage should be as small as possible. In the mathematical form this condition is written as

$$h^* < l_m \ . \tag{21}$$

(ii) Condition of *representativity or macroscopicity*. On the other hand, such a small and (approximately) homogeneous sample should nevertheless be sufficiently large to have a macroscopic size, i.e., contain an enormous number of aggregates. Then, the small sample (in the meaning of Eq.(21)) can still be considered a representative elementary volume of the soil matrix (Bear and Bachmat, 1990). In the mathematical form this condition is written as

$$l_{min} \ll h^* \ . \tag{22}$$

Critical size, $h^*$ plays the part of the *boundary* between "small" samples of size, $h$ meeting the condition, $l_{min} \ll h < h^*$ and "large" ones of size $h > h^*$. With high probability the small samples are only subject to deformation at shrinkage. The large ones are also subject to cracking. The larger $h$ is (at $h > h^*$), the larger the crack volume at a given water content. To estimate $l_{min}$, $l_m$, and then $h^*$ we should first specify the $l_{min}$ and $l_m$ concepts. We are interested in the mean distances, $l_{min}$ and $l_m$ between the smallest and largest aggregates, respectively. In both cases we imply the aggregates of sizes in a certain *small* range $\Delta X$ adjoining to the minimum ($X_{min}$) or maximum ($X_m$) aggregate sizes, respectively, but not the aggregates of exact size, $X_{min}$ or $X_m$ (since the number of aggregates of size, $X_{min}$ or $X_m$ exactly, is equal to zero). With that the small range, $\Delta X$ should be *physically small*. That is, $\Delta X$ should simultaneously comply with the following conditions: (i) $\Delta X$ is maximally small compared to $X_m$ ($\Delta X \ll X_m$); but at the same time (ii) $\Delta X$ exceeds the maximum size, 0.05 mm of the smallest sub-aggregate particles (clay particles and silt grains); and (iii) $\Delta X$ exceeds the minimum aggregate size, $X_{min}$. The first condition is obvious. The two last conditions should be fulfilled since $\Delta X$ is the range of the aggregate size axis. Accounting for that usually $X_{min} \ll X_m$ one comes to the $\Delta X$ estimate as

$\Delta X = \max(0.05\text{mm}, X_{min}) \ll X_m$ (23)

Thus, $l_{min}$ is understood as the mean distance between aggregates of size $X$ in the range $X_{min} \leq X \leq X_{min}+\Delta X$ adjoining to $X_{min}$ ($\Delta X$ from Eq.(23)) with the mean aggregate size $X=X_{min}+\Delta X/2$. Similarly, $l_m$ is understood as the mean distance between aggregates of size $X$ in the range, $X_m-\Delta X \leq X \leq X_m$ adjoining to $X_m$ with the mean aggregate size $X=X_m-\Delta X/2$. The mean distances, $l_{min}$ and $l_m$ are determined by cumulative aggregate-size distribution at maximum swelling, $F(X, X_{min}, X_m, P_h)$ from the intersecting surfaces approach (Chertkov, 2005b) as follows. (i) The *volume* of aggregates of size $X$ in the small range $\Delta X$ (Eq.(23)) per unit volume of all aggregates is $(dF/dX)\Delta X$ (by definition of the $F$ distribution ). (ii) The *number* of such aggregates per unit volume of all aggregates is $(1/X^3)(dF/dX)\Delta X$. (iii) The *volume of all aggregates* per *one* such aggregate of size $X$ is the reciprocal: $X^3/[(dF/dX)\Delta X]$. (iv) The *soil volume* per *one* such aggregate of size $X$ is $(1-P_h)^{-1}X^3/[(dF/dX)\Delta X]$ ($P_h$ being the inter-aggregate porosity at maximum swelling). (v) The mean distance $l$ in the soil between aggregates of size $X$ in the small range $\Delta X$ is $l=(1-P_h)^{-1/3}X/[(dF/dX)\Delta X]^{1/3}$.
(vi) Thus, by definition of $l_{min}$ and $l_m$ and accounting for $\Delta X$ (Eq.(23))

$l_{min}=(1-P_h)^{-1/3}(X_{min}+\Delta X/2)/[(dF/dX)_{|X=X_{min}+\Delta X/2}\Delta X]^{1/3}$ , (24a)

$l_m=(1-P_h)^{-1/3}(X_m-\Delta X/2)/[(dF/dX)_{|X=X_m-\Delta X/2}\Delta X]^{1/3}$ . (24b)

Now one should estimate $h^*$ from $l_{min}$ and $l_m$ (Eq.(24)). It is obvious that the physical conditions from Eq.(20) do not determine $h^*$ in a single-valued manner, but only by the order of magnitude as $10l_{min} \sim 0.1l_m \cong < h^* < l_m$ (since $l_m/l_{min} \sim 10^2$). For this reason we, first, *define* the *rough approximation*, $h^*_o$ of the $h^*$ value as

$h^*_o = (l_{min}l_m)^{1/2}$ . (25)

In the task of the $h^*$ presentation there are two independent parameters with the dimension of length, e.g., $X_{min}$ and $X_m$, or $l_{min}$ and $l_m$, and others. However, the most convenient choice of the parameter pair is $h^*_o$ and $X_m$ since $h^* \sim h^*_o$ and $X_m/h^*_o \sim 0.1$. One can then write $h^*$ from the dimension considerations as $h^*=f(X_m/h^*_o)h^*_o$. Here, from the geometrical considerations, the $f$ function should, in fact, depend on the ratio, $X_m^3/h^{*3}_o$ of the maximum aggregate volume, $X_m^3$ to the $h^{*3}_o$ volume because the samples under consideration have three close sizes. Further, since $f \to 0$ at $X_m \to 0$ (from the physical meaning of $X_m$ and $h^*$) and since $(X_m/h^*_o)^3 \sim 10^{-3}$ the $f$ function is proportional to $(X_m/h^*_o)^3$. Since $h^* \sim h^*_o$ (and $(X_m/h^*_o)^3 \sim 10^{-3}$) the proportionality coefficient is close to $10^3$ (*assumption* 2). Thus, $h^*$ for any soil is presented as

$$h^* = 10^3 (X_m/h^*_o)^3 h^*_o \ . \tag{26a}$$

This theoretical coefficient ($10^3$) can be specified using the relevant data on a sufficiently large number of different soils. The substantiation of the $h^*$ presentation from Eqs.(24)-(26a) using available data will be considered in Section 11.3.

To obtain the illustrative numerical estimates of $l_{min}$, $l_m$, $h^*_o$, and $h^*$ for the real $F$ distributions (see ISA approach (Chertkov, 2005b)), first, we replace $(1-P_h)^{-1/3}$ in Eqs.(24a) and (24b) with unity since usually $P_h \cong 0\text{-}0.15$ and $1 < (1-P_h)^{-1/3} < 1.05$. Second, we take the reasonable values of $X_{min} \cong 20\text{-}70\mu m$ and $X_m \cong 2\text{-}5mm$. Table 1 shows the $l_{min}$ and $l_m$ estimates for the possible pairs of boundary values of $X_{min}$ and $X_m$. The $h^*_o$ (Eq.(25)) and $h^*$ (Eq.(26a)) estimates are also shown in Table 1, according to which $h^*$ is within the limits of the first centimeters. For a particular soil the estimates of $l_{min}$ (Eq.(24a)), $l_m$ (Eq.(24b)), and $h^*$ (Eqs.(25) and (26a)) can be specified. According to Table 1 in the aggregated soil the relative mean distance between the smallest aggregates, $l_{min}/X_{min}$ can be both more and less than that between the largest aggregates, $l_m/X_m$. It should be noted that in a number of experiments on soil shrinkage (Crescimanno and Provenzano, 1999; Braudeau et al., 2005; Boivin et al., 2006; Cornelis et al., 2006) the sample sizes without observable cracking, meet the condition $h < h^* \cong 2\text{-}5cm$. Conversely, in the samples of size $h > h^* \cong 2\text{-}5cm$ cracks were observed (Crescimanno and Provenzano, 1999; Cornelis et al., 2006). Thus, the above theoretical estimate of $h^*$ (by the order of magnitude) is in agreement with the available data. Finally, we should indicate the applicability condition of the above estimating of the critical sample size, $h^*$ according to Eqs.(20), (23)-(26a). With the decrease in clay content soil shrinkage weakens. At sufficiently weak shrinkage even the maximum possible shrinkage stresses become less than the soil strength. Hence, cracking does not occur, even in large samples of soils with such weak shrinkage. It means that with a decrease in clay content, the critical sample size, $h^*$ grows (depending on the soil texture and structure) and, eventually, can exceed the $l_m$ value (see Eq.(20)) (for instance, it is clear that for rigid soils, formally, $h^* \to \infty$). For this reason one can obtain the applicability condition of the above $h^*$ estimate, substituting for $h^*$ in the condition, $h^* < l_m$ (Eq.(20)) its expression from Eq.(26a) as

$$10^3 (X_m/h^*_o)^3 h^*_o < l_m \ . \tag{26b}$$

The violation of this condition for a particular soil will mean its sufficiently weak shrinkage (vs. its strength properties) and inapplicability of the above way of estimating $h^*$. It is clear, however, that for clay soils this condition should be fulfilled.

### 7. The crack factors for a sample ($q_s$) and layer ($q_l$)

The objective of this section is to derive the crack factor expressions in sample and layer cases. The increment $dU'(w')$ of the intra-aggregate matrix volume depends on clay type and content through the increment $dU'_{cp}(w')$ of the clay pore volume and lacunar factor, $k$ (Eqs.(10) and (19)). For this reason it is natural to assume that at a given increment, $dU'(w')$ both the increments of soil ($dY(w')$) and of crack ($dU_{cr}(w')$) volume (Eqs.(12) and (13)) do not depend on the clay type and content in the soil. In other words, $dY(w')$ and $dU_{cr}(w')$ can only depend on clay content and type through $dU'(w')$ (Eq.(10)), but not through the crack factor, $q$. That is the latter only depends on sample shape and sizes. We consider the simplest case of a sample that can be characterized by one size $h$ (i.e., with three finite close sizes). It can be the sample of the approximately cubic or cylindrical shape (close in height and diameter). We designate the crack factor of such a sample as $q_s$. The cracks develop between aggregates from the initial structural (inter-aggregate) pores (Fig.1). For this reason and according to its physical meaning, the $q_s$ value can only depend (except for the sample size, $h$) on the characteristics of the aggregate-size distribution, such as $X_{min}$, $X_m$, and $P_h$ (see previous section). These characteristics enter the expression for the critical sample size, $h^*$ (Eqs.(24)-(26a)) that plays the part of the *internal soil scale*. In addition, $q_s$ is the dimensionless value. For this reason it is assumed that the crack factor of a soil sample, as a function of the sample size, $h$ is a universal function $q_s(h/h^*)$ of the $h/h^*$ ratio (the *sample scale parameter*).

The $q_s(h/h^*)$ dependence can be constructed as follows. In the small samples ($h<h^*$) cracks do not appear at shrinkage (see section 6). It means that at $h<h^*$ in Eq.(12) $dU_{cr}(w')=0$. Hence, at $0<h/h^*<1$ $q_s(h/h^*)=0$ (Fig.4) (see Eq.(12) where $dU'(w')\neq 0$). With an *imagined* sample size increase in the range including $h=h^*$ and at any possible *fixed* water content, the probability of crack appearance and opening smoothly increases starting from zero value at $h=h^*$. In other words the $q_s(h/h^*)$ dependence should be smooth at $h/h^*=1$; that is, $q_s(h/h^*=1)=0$ and $dq_s/d(h/h^*)_{|h/h^*=1}=0$ (Fig.4). Therefore, in the sufficiently small range $1\leq h/h^*<1+\varepsilon$, adjoining to $h/h^*=1$, $q_s$ can be presented as $q_s\cong b_1(h/h^*-1)^2$ where $b_1$ is a universal constant (i.e., one that relates to any shrinking and cracking soil). Such a presentation is just the first non-disappearing term of the expansion of $q_s$ in powers of $(h/h^*-1)$ in the vicinity of $h/h^*=1$. In connection with the $q_s$ behavior at large $h/h^*$ values, it is clear that at $h/h^*\to\infty$, the shrinkage of the intra-aggregate matrix ($dU'(w')$ in Eqs.(12) and (13)) is *totally* transformed to crack volume increase ($dU_{cr}(w')$) (indeed, at very large soil volume it is meaningless to speak about the shrinkage of the soil volume as a whole, i.e., $dY=0$ in Eq.(13) at $h/h^*\to\infty$). Hence, $q_s(h/h^*\to\infty)\to 1$ (see Eq.(12) and Fig.4). Therefore, at sufficiently large $h/h^*$, $q_s$ can be presented as $q_s\cong 1-b_2/(h/h^*-1)$ where $b_2$ is also a universal constant. Such a presentation corresponds to the first terms of the expansion of $q_s$ in powers of $(h/h^*-1)^{-1}$ at large $h/h^*$. Figure 4 shows the qualitative view of the corresponding $q_s(h/h^*)$ dependence throughout the range, $0<h/h^*<\infty$. Accounting for this qualitative view and the above quantitative behavior of $q_s$ at $0<h/h^*<1$, in the vicinity of $h/h^*=1$, and at $h/h^*>>1$, one can suggest for $q_s(h/h^*)$ (Fig.4) the following quantitative presentation in the case of the sample with three close sizes $\sim h$ (*assumption* 3)

$$q_s(h/h^*)=0, \qquad 0<h/h^*\leq 1 \tag{27a}$$

$$q_s(h/h^*)=b_1(h/h^*-1)^2, \qquad 1\leq h/h^*\leq 1+\delta \tag{27b}$$

$$q_s(h/h^*)=1-b_2/(h/h^*-1), \qquad h/h^*\geq 1+\delta \tag{27c}$$

This presentation is obtained by continuing of the $q_s$ presentations at $h/h^*$ close to $h/h^*=1$ and at $h/h^*>>1$ beyond the limits of their formal applicability, up to a "sewing" point, $h/h^*=1+\delta$ (Fig.4). The smoothness conditions of $q_s(h/h^*)$ in the "sewing" point permit one to express $\delta$ and $b_2$ through $b_1$ as

$$\delta=(1/(3b_1))^{1/2} \quad \text{and} \quad b_2=2\delta/3=(2/3)(1/(3b_1))^{1/2} \ . \tag{27d}$$

One can theoretically estimate the $b_1$, $b_2$, and $\delta$ values from the obvious condition $q_{s|h/h^*>>1}\cong 1$ (Fig.4) or, more specifically, $q_{s|h/h^*\cong 100}\cong 0.99$. Then, Eq.(27c) gives $b_2\cong 1$, and from Eq.(27d) $b_1\cong 0.15$ and $\delta\cong 1.5$. More accurate estimates of the universal $b_1$, $b_2$, and δ values in the $q_s$ presentation (Eq.(27a)-(27c)) can be found from the relevant experimental data on the sufficiently large number of different soils.

Now based on the expression for $q_s$ (Eq.(27)) we can construct the single-valued expression for the layer crack factor, $q_l$ of thickness $h$. Similar to $q_s$ the $q_l$ factor is a function of the $h/h^*$ ratio, similar to $q_s(h/h^*)$ at $h/h^* \to \infty$ $q_l \to 1$ (Fig.4). However, unlike $q_s(h/h^*)$, $q_l(h/h^*)>0$ in the $0<h/h^* \leq 1$ range (Fig.4) since two layer sizes are always much more than $h^*$. In addition, $q_l(h/h^*)$ in this range cannot be less than $q_s(h'/h^*)$ in the corresponding points, $h'/h_*=h/h^*+1$ of the $1 \leq h'/h^* \leq 2$ range (Fig.4), because at a given $h$ two layer sizes are much more than corresponding $h$ sizes of the cube. On the other hand, in the case of the sample at $h'/h^* \cong 1$, all three its sizes grow, and in the case of the layer at $h/h^* \cong 0$, only one size (layer thickness) increases. For this reason $q_l(h/h^*)$ should grow in the vicinity of $h/h^* \cong 0$ no quicker than $q_s(h'/h^*)$ in the vicinity of $h'/h^* \cong 1$. Thus, we come to the conclusion that $q_l(h/h^*)=q_s(h'/h^*)$ at $0<h/h^*<\infty$ and $h'=h+h^*$. Hence, accounting for Eq.(27) one can write $q_l$ as (Fig.4)

$$q_l(h/h^*)=b_1(h/h^*)^2, \qquad 0 \leq h/h^* \leq \delta \tag{28a}$$

$$q_l(h/h^*)=1-b_2/(h/h^*), \qquad h/h^* \geq \delta \tag{28b}$$

with the same $b_1$, $b_2$, and δ. The experimental validation of *assumption* 3, i.e., $q_s$ factor dependence (and hence, $q_l$ factor dependence also) with indicated $b_1$, $b_2$, and δ on the $h/h^*$ ratio (the scale parameter) is regarded in Section 11.3.

**8. Specific soil and crack volumes as functions of initial sample/layer size**

If we know $q_s(h/h^*)$ (Eq.(27)), Eqs.(14) and (15) at $q=q_s$ permit us to find the specific crack ($U_{cr\,s}$) and soil ($Y_s$) volume for the sample case (the "s" index) to be

$$U_{cr\,s}(w',h/h^*)=q_s(h/h^*)\,(U_h-U(w))/K+U_s, \qquad 0 \leq w' \leq w'_h, \quad 0 \leq w \leq w_h, \tag{29}$$

$$Y_s(w',h/h^*)=(1-q_s(h/h^*))U(w)/K+q_s(h/h^*)U_h/K+U_s+U_i, \qquad 0 \leq w' \leq w'_h, \quad 0 \leq w \leq w_h. \tag{30}$$

Similar expressions can be written for the specific crack ($U_{cr\,l}$) and soil ($Y_l$) volumes of the layer case (the "l" index) from Eqs.(28), (14) and (15) at $h/h^*>0$ as

$$U_{cr\,l}(w',h/h^*)=q_l(h/h^*)\,(U_h-U(w))/K+U_s, \qquad 0 \leq w' \leq w'_h, \quad 0 \leq w \leq w_h, \tag{31}$$

$$Y_l(w',h/h^*)=(1-q_l(h/h^*))U(w)/K+q_l(h/h^*)U_h/K+U_s+U_i, \qquad 0 \leq w' \leq w'_h, \quad 0 \leq w \leq w_h, \tag{32}$$

(for $w'$, $w$, and $W$ see Eqs.(6) and (1)). It is worth emphasizing the following consequence of Eqs.(27)-(28) and Fig.4: $q_l(h/h^*) \cong q_s(h/h^*)$ and is close to unity at $h/h^* > 10 - 15$. That is, the shrinkage curves of a sample (Eq.(30)) and layer (Eq.(32)) at such $h$ values are close. The same relates to the specific crack volumes in the sample (Eq.(29)) and layer (Eq.(31)). Therefore, to determine the layer shrinkage curve ($Y_l$) at layer thickness $h/h^*>10 - 15$ one can conduct the estimates or measurements using the samples of the size $h/h^*>10-15$. Thus, Eqs.(27)-(32) determine the effect of the initial sample/layer size on the specific crack, sample, and layer volume. Note, that by definition of the initial crack volume ($U_{cr\,s}(w'_h)=U_{cr\,l}(w'_h)=U_s$) and reference shrinkage curve, $Y_r(W)$ (Chertkov, 2007a, 2007c), the latter is found to be

$Y_r = Y_s - U_{cr\,s} + U_s = Y_l - U_{cr\,l} + U_s$ ,  (33)

and does not depend on sample size and shape.

**9. Physical parameters determining the soil shrinkage and cracking**

The specific soil volume, $Y(W, h/h^*)$ and specific crack volume, $U_{cr}(W, h/h^*)$ for the sample and layer cases are determined by: (i) the initial sample/layer size, $h$; (ii) the critical sample size, $h^*$; and (iii) the basic dependence, $Y_r(W)$ of the reference shrinkage curve that was considered in Chertkov (2007a, 2007b) including the eight physical parameters determining $Y_r(W)$: oven-dried specific soil volume ($Y_{rz}$), maximum swelling water content ($W_h$), mean solid density ($\rho_s$); soil clay content ($c$); oven-dried structural porosity ($P_z$), ratio of aggregate solid mass to solid mass of intra-aggregate matrix ($K$), the lacunar factor ($k$), and lacunar pore volume in the oven-dried state ($U_{lpz}$). The general prediction algorithm of the reference shrinkage curve, $Y_r(W)$ shows that input parameters ($Y_{rz}$, $W_h$, $\rho_s$, $c$, $P_z$, $K$, $k$, and $U_{lpz}$) permit one to find two fundamental clay matrix parameters (for the clay contributing to the soil), the relative volumes of clay solids, $v_s$ and the oven-dried clay, $v_z$. These parameters can also be found independently of the above input parameters from the measurements of the clay (Chertkov, 2000b, 2003, 2005a). Thus, instead of $Y_{rz}$ and $W_h$, $v_s$ and $v_z$, if they are available, can be used as independent and more fundamental input parameters connected with soil structure. Such new parameters as $P_z$, $K$, $k$, and $U_{lpz}$ are considered in more detail below because they can be expressed through more customary ones or reduced to elementary parameters of soil structure and texture.

*9.1. Parameter $P_z$*

$P_z$ can be directly determined from the comparison between the experimental aggregate-size distribution and distribution, $F(X, X_{minz}, X_{mz}, P_z)$ in the framework of the intersecting-surfaces approach (ISA) (Chertkov, 2005b) as

$F(X, X_{minz}, X_{mz}, P_z) \equiv F(\eta, P_z) = (1 - P_z^{I_o(\eta)/8.4})/(1-P_z)$ ,  (34)

($X_{minz}$ and $X_{mz}$ are the minimum and maximum aggregate sizes at $W \leq W_z$; Fig.2);

$\eta \equiv (X - X_{minz})/(X_{mz} - X_{minz})$,   $X_{minz} \leq X \leq X_{mz}$ ,  (35)

$I_o(\eta) = \ln(6)\,(4\eta)^4 \exp(-4\eta)$ .  (36)

Still another way is also possible. If the reference shrinkage curve has a horizontal part at $W > W_h$, (Fig.2), its size along the water content axis, $\Delta W = \rho_w U_s$ determines the specific volume of structural pores, $U_s$ (Fig.2). $P_z$ can be found to be $P_z = U_s/Y_{rz}$. Note also that $P_z$ can be replaced with the structural porosity at $W = W_h$, $P_h = U_s/Y_h$ (see $Y_h$ in Fig.2) since they are connected (Chertkov, 2007b, 2008b).

*9.2. Parameter $k$*

To find $c_*$ in the calculation of the lacunar factor, $k(c/c_*)$ (section 5) one needs to know $v_s$, $v_z$, and the porosity, $p$ of contributive silt and sand grains when they are in the state of (*imagined*) contact (Chertkov, 2007a). Analogously to structural porosity, $P_z$, $p$ can be directly estimated from the comparison between the experimental size distribution of contributive silt and sand grains and the distribution, $f(x, x_{min}, x_m, p)$ ($x_{min}$ and $x_m$ are minimum and maximum grain size) in the frame of the intersecting-surfaces approach (ISA) (Chertkov, 2005b) (cf. Eqs.(34)-(36) with replacement: $F \to f$, $X \to x$; $P_z \to p$; $X_{minz} \to x_{min}$; $X_{mz} \to x_m$) as

$$f(x, x_{\min}, x_{\text{m}}, p) \equiv f(\eta, p) = (1-p^{I_o(\eta)/8.4})/(1-p) , \tag{37}$$

$$\eta \equiv (x-x_{\min})/(x_{\text{m}}-x_{\min}), \quad x_{\min} \leq x \leq x_{\text{m}} , \tag{38}$$

and $I_o(\eta)$ from Eq.(36). Still another way to estimate $p$ is as follows. One finds the $\eta$ value corresponding to the maximum silt grains, $x=x_{\text{silt}} \cong 50\mu$m at $x_{\min} \cong 2\mu$m as

$$\eta_{\text{silt}} = 48/(x_{\text{m}}-2) . \tag{39}$$

Then, the fraction of the silt grains among all silt and sand grains, $f_{\text{silt}}$ is found to be

$$f_{\text{silt}} = (1-p^{I_o(\eta_{\text{silt}})/8.4})/(1-p) . \tag{40}$$

Equation (40) determines the $p$ value at a given $f_{\text{silt}}$ and maximal sand grain size, $x_{\text{m}}$.

*9.3. Parameter K*

The calculation of the $K$ ratio was considered in Chertkov (2008b). It is conducted based on the above indicated data set, $Y_{\text{rz}}$, $W_{\text{h}}$, $\rho_{\text{s}}$, $c$, $P_{\text{z}}$, $k$, and $U_{\text{lpz}}$ (but without $K$) and two parameters characterizing the soil structure, $X_{\text{mz}}$ (the maximum aggregate size in the oven-dried state) and texture, $x_{\text{n}}$ (the mean size of soil solids: clay particles, silt, and sand grains; $x_{\text{n}} \cong X_{\text{minz}}$). $X_{\text{mz}}$ and $x_{\text{n}}$ are found from the granulometric analysis of aggregates and soil solids, respectively.

*9.4. Parameter $U_{\text{lpz}}$*

The specific lacunar pore volume in the oven-dried soil, $U_{\text{lpz}}$ can be replaced by a similar value at maximum swelling, $U_{\text{lph}}$ (Fig.2) (Chertkov, 2007c, 2010b) as $U_{\text{lph}} = U_{\text{lpz}} - k(1-u_{\text{S}})(v_{\text{h}}-v_{\text{z}})/(\rho_{\text{s}} u_{\text{S}})$. In turn, $U_{\text{lph}}$ is determined through the displacement of the true saturation line relative to the pseudo saturation line along the water content axis (Fig.2) as $U_{\text{lph}} = (W_{\text{h}}^*-W_{\text{h}})/\rho_{\text{w}} = (W_{\text{m}}^*-W_{\text{m}})/\rho_{\text{w}}$ where $W_{\text{m}} = W_{\text{h}} + U_{\text{S}}\rho_{\text{w}}$, and $W_{\text{h}}^*$ is determined from $U_{\text{ah}} = 1/\rho_{\text{s}} + W_{\text{h}}^*/\rho_{\text{w}}$ because point ($W_{\text{h}}^*$, $U_{\text{ah}}$) should lie on the true saturation line (Fig.2), and $W_{\text{m}}^*$ is determined from $Y_{\text{h}} = 1/\rho_{\text{s}} + W_{\text{m}}^*/\rho_{\text{w}}$ because the ($W_{\text{m}}^*$, $Y_{\text{h}}$) point should also lie on the true saturation line (Fig.2).

Thus, at the calculations of $Y_{\text{r}}(W)$ with the above indicated data set ($Y_{\text{rz}}$, $W_{\text{h}}$, $\rho_{\text{s}}$, $c$, $P_{\text{z}}$, $K$, $k$, $U_{\text{lpz}}$ or $v_{\text{s}}$, $v_{\text{z}}$, $\rho_{\text{s}}$, $c$, $P_{\text{h}}$, $K$, $k$, $U_{\text{lph}}$), one can use the minimum and maximum aggregate sizes, $X_{\text{minz}}$ ($\cong x_{\text{n}}$) and $X_{\text{mz}}$, as the basic parameters instead of $K$, porosity $p$ instead of $k$, and the specific soil volume at maximum swelling, $Y_{\text{h}}$ instead of $U_{\text{lpz}}$ (or $U_{\text{lph}}$). The parameters, $X_{\text{minz}}$, $X_{\text{mz}}$, $p$, and $Y_{\text{h}}$ are obviously customary values.

Note that in order to calculate the critical sample size $h^*$ one needs parameters $P_{\text{z}}$, $X_{\text{minz}}$, $X_{\text{mz}}$ (Eqs.(23)-(26a)) that have already entered the final data set. Thus, this final data set includes ten *physical* parameters: $v_{\text{s}}$, $v_{\text{z}}$ (or $Y_{\text{rz}}$, $W_{\text{h}}$), $\rho_{\text{s}}$, $c$, $P_{\text{z}}$ (or $P_{\text{h}}$), $Y_{\text{h}}$, $p$, $X_{\text{minz}}$ ($\cong x_{\text{n}}$), $X_{\text{mz}}$, $h$. In the case of sufficiently high clay content, $c \geq c_*$ porosity $p$ is not needed because $k=0$ and $Y_{\text{h}}$ is not needed because $U_{\text{lpz}} = U_{\text{lph}} = 0$ (Chertkov, 2007a). That is, at $c \geq c_*$ one needs eight parameters: $v_{\text{s}}$, $v_{\text{z}}$, $\rho_{\text{s}}$, $c$, $P_{\text{z}}$, $X_{\text{minz}}$, $X_{\text{mz}}$, $h$. When using the presentation of the shrinkage curve in coordinates of moisture ratio – void ratio instead of $W - Y$, we do not need the $\rho_{\text{s}}$ parameter, and the input data set of the physical parameters is reduced to nine and seven at $c<c_*$ and $c \geq c_*$, respectively. The ten parameters can be divided into three groups by their physical meaning: (i) *intra-aggregate* parameters that characterize the internal structure of aggregates, $v_{\text{s}}$, $v_{\text{z}}$, $\rho_{\text{s}}$, $c$, $p$; (ii) *inter-aggregate* parameters that characterize the soil as a whole, accounting for its aggregate structure (aggregate-size distribution), $P_{\text{z}}$, $Y_{\text{h}}$, $X_{\text{minz}}$, $X_{\text{mz}}$; and (iii) *initial* *sample size* or *layer thickness*, h of a given soil (h can be of any value; the only limitation being the homogeneous water content within the sample or layer volume) at the *chosen* sample shape (approximately cubic or cylinder samples at $W=W_h$).

**10. Data used**

Many works containing different experimental data on soil shrinkage, cracking, and shrinkage curves are available in the literature. However, these works all have different aims. For this reason the data are very limited that are relevant to the validation of the approach which permits the physical prediction of the soil shrinkage curve. We could not find such data directly relating to the soil layer case. Fortunately, the results for this case in a single-valued manner follow from those for the sample case (Section 7). In turn, the data for the sample case should meet a number of conditions that flow out of the approach. According to the first condition, the measurements of the soil shrinkage curve should be conducted on samples of, at least, two different sizes. We could only find four such works (Yule and Ritchie, 1980a, 1980b; Crescimanno and Provenzano, 1999; Cornelis et al., 2006). Further, the h sizes of smaller and larger samples should meet the additional condition: $h<h^*$ and $h>h^*$, respectively. Accounting for the preliminary estimates of the critical sample size, $h^*\cong 2$-5cm (Section 6; Table 1), the sizes of smaller samples should be $h<2$-5cm and those of larger samples, $h>5$cm. "Small" samples from Yule and Ritchie (1980a) do not meet this condition (their size was ~10cm). In addition, for dependable checking it is very desirable that the larger sample size be at least twice as large as the smaller one. Samples from Cornelis et al. (2006) do not meet this condition (the smaller samples are ~3-3.6cm and larger ones are ~4.6cm). Finally, according to the obvious condition, the aggregate structure of the small and large soil samples should be similar (i.e., the aggregate-size distributions should be similar). Samples from Cornelis et al. (2006) do not meet this condition either since the smaller samples were prepared with preliminary soil destruction and sieving. As a result, only data relating to the four soils from Crescimanno and Provenzano (1999) were more or less suitable for our aims (these authors consider more soils, but only give the shrinkage curve data for the four soils). The primary experimental data that we used, included the *numerical* data on: (i) clay (c), silt ($s_1$), and sand content ($s_2$) (Table 2); (ii) volume of small samples (clods) - 20-30cm$^3$; (iii) sizes of large samples (cores) - height, 11.5cm and diameter, 8.5cm; and *graphical* data presented by (i) experimental points in Figs.5-8 for the clod shrinkage curve (white circles) and core shrinkage curve (black circles) as well as (ii) position of the saturation line in Figs.5-8.

**11. Data analysis in the sample case**

*11.1. Estimating some soil structure characteristics from small-sample data*

The integrated approach permits the physical prediction (i.e., without fitting) of both the reference shrinkage curve, $Y_r(W)$ and shrinkage curve of large samples with cracks, $Y_s(W)$. However, for such a prediction one needs to know a number of soil structure characteristics (section 9). Not all of them are available in Crescimanno and Provenzano (1999). To compensate for this drawback we used in this section the experimental data on small-sample shrinkage curves in Figs.5-8 (white circles).

Table 2, in addition to the primary data, shows all the input parameter values that are necessary for the prediction, except for K and k. The $Y_{rz}$ value in Table 2 for each soil is the average on the experimental values at $W=0$ (the white circles in Figs.5-8); $W_h$ is the maximum experimental W value (the white circles in Figs.5-8); the mean solid density, $\rho_s$ is determined by the intersection point position of the saturation line and specific volume axis in Figs.5-8; clay content, c enters the primary data (Table 2); the structural porosity, $P_z=0$ for all soils under consideration since the experimental

shrinkage curves (Figs.5-8, the white circles) do not have the horizontal part at the intersection with the saturation line (see section 9; cf. Fig.2 at $U_s$=0); the lacunar pore volume at maximum swelling, $U_{lph}$=0 for all four soils (cf. Fig.2) since the experimental points at $W=W_h$ for the soils are on the saturation line (see Figs.5-8, the white circles). Note that the above input parameters (Table 2) were estimated without any using of the data (white circles) at $0<W<W_h$.

Using recent results (Chertkov, 2008b, 2010b), we intend to estimate the aggregate/intra-aggregate mass ratio, $K$ and the lacunar factor, $k$ for the above four soils based on the more fundamental parameters of the soil texture and structure (cf. section 9). However, it is these texture and structure parameters that are absent from the data on the above four soils. Nevertheless, let us assume for a moment that the maximum aggregate size in the oven-dried state, $X_{mz}$ and maximum sand grain size, $x_m$ are known. The general algorithm of the $K$ and $k$ estimation in such a case is given in the Appendix. The algorithm relies on the primary data and input parameters from Table 2 as well as the $X_{mz}$ and $x_m$ values, and combines the results from Chertkov (2007a, 2007c, 2008b, 2010b). Then, according to Chertkov (2007a, 2007c) the input parameters (Table 2) and the $K$ and $k$ values found allow one to physically predict the reference shrinkage curve. However, since $X_{mz}$ and $x_m$ are in fact not known, we modified the indicated prediction procedure of the reference shrinkage curves of the four soils as follows. For a given soil we used the prediction from Appendix for a set of possible pairs of the $X_{mz}$ and $x_m$ values (see below) and found the corresponding set of the $K$ and $k$ pair values and the set of corresponding reference shrinkage curves. Then, based on some criteria (see below) we selected for each soil a much narrower range of $X_{mz}$ and $x_m$ values, a corresponding much narrower range of the $K$ and $k$ values, and a single-valued reference shrinkage curve. In other words, using (i) the general algorithm of the $K$ and $k$ estimation (see Appendix); (ii) prediction of the reference shrinkage curve at given $K$ and $k$ (Chertkov, 2007a, 2007c); and (iii) the data on the reference shrinkage curves (Crescimanno and Provenzano, 1999) (Figs.5-8, the white circles), we showed that for the $X_{mz}$ and $x_m$ values and together with them for $K$ and $k$ of each soil one can find such small ranges that the inaccuracy of $K$ and $k$ practically does not influence the corresponding reference shrinkage curve. Below, this analysis and its results are described in more detail in a number of points.

1. The maximum sand grain size, $x_m$ and maximum aggregate size in the oven-dried state, $X_{mz}$ should satisfy two conditions (0.05mm is the maximum silt grain size) as

$$x_m > 0.05\text{mm}, \qquad X_{mz} > x_m \qquad\qquad (41)$$

2. The direct calculation for each soil using the algorithm from the Appendix shows that at sufficiently small $x_m$ in a range, $0.05\text{mm}<x_m<x_{m\ min}$ there is no solution of Eq.(A14) (see the Appendix) for the relative volume $v_z$ of the contributive clay in the oven-dried state, and there are no corresponding values of the critical clay content, $c_*$ and lacunar factor, $k$ at any possible $X_{mz}$ value from Eq.(41). The solution for $v_z$ (and corresponding $c_*$ and $k$ values) only appear at $x_m > x_{m\ min}$ and at the sufficiently large $X_{mz}$ values from Eq.(41) (for these $X_{mz}$ see point 6 below). That is, for each soil at given primary data and input parameters from Table 2, only the $x_m$ values from the range, $x_m > x_{m\ min} > 0.05\text{mm}$ can be physically realized. This lower border, $x_{m\ min}$ of the physically possible $x_m$ range for each soil is indicated in Table 3.

3. A similar direct calculation for each soil shows that with the $x_m$ increase in the above physically realized range (at given data and parameters from Table 2) and at sufficiently large $X_{mz}$ from Eq.(41), the $p$ porosity of the contributive silt and sand grains in the *imagined* contact state (see section 9 and Appendix) decreases, and at sufficiently large $x_m$ becomes unreasonably small ($p<0.1$) for a silt-sand mixture. Thus, the natural condition, $p>0.1$ determines the upper border, $x_m=x_{m\ max}$ of the physically possible $x_m$ range for each soil from Table 2. The upper border, $x_{m\ max}$ of the physically possible $x_m$ range for each soil is also indicated in Table 3.

4. The statements in points 2 and 3 determine for each soil the relatively narrow range, $0.05\text{mm}<x_{m\ min}<x_m<x_{m\ max}$ (Table 3) and corresponding range for the $p$ porosity (Table 3). This narrow range of $x_m$ is only determined by general physical considerations (see points 1-3) as well as data and parameters from Table 2, but not the experimental points in Figs.1-4 (white circles) at $0<W<W_h$. Table 3 also shows the range of the mean soil-solids size, $x_n$ (Eq.(A6)). Note that $x_n \cong X_{minz}$ (Chertkov, 2008b).

5. If the $x_m$ value is in a small part of the physical range, $x_{m\ min}<x_m<x_{m\ max}$, namely in the small vicinity of $x_{m\ min}$, Eq.(A14) from the Appendix (at data and parameters from Table 2) provides two solutions for $v_z$ and two corresponding values for $c_*$ and $k$. One solution corresponds to $k=0$ and $c\geq c_*$, and another to $0<k<1$ and $c<c_*$. Of course, for each soil only one of two possibilities is realized (see point 6). Note that the $p$ porosity (Table 3) only influences the lacunar factor, $k$ if $0<k<1$ (Chertkov, 2007c).

6. To estimate for each soil the $X_{mz}$ values at $x_{m\ min}<x_m<x_{m\ max}$ we used the data from Figs.5-8 (white circles) at $0<W<W_h$ and the least-square criterion as follows. For a value set of $X_{mz}>x_m$ and $x_{m\ min}<x_m<x_{m\ max}$ we estimated $K$ (see Appendix) at data and parameters from Table 2. Then for each pair, $K, k$ ($k$ was found earlier) we found the reference shrinkage curve, $Y_r(W)$ as in Chertkov (2007a, 2007c) and the sum of squares, $\Sigma$ of the differences between the experimental values $Y_{re}$ (white circles in Figs.5-8) and the corresponding theoretical $Y_r(W_e)$ values. The variation of $X_{mz}>x_m$ at $x_{m\ min}<x_m<x_{m\ max}$ allowed for the estimation of some $X_{mz}$ range (see Table 3) and the corresponding range of $K$ (Table 3) where $\Sigma$ stays constant and minimum, $\Sigma=\Sigma_{min}$. The corresponding goodness of fit, $r^2_{Yr} \cong 0.95-0.99$ (see the captions of Figs.5-8). In particular, for all $x_m$ for which both $k=0$ and $k>0$ are possible (see the above point 5), the case of $k>0$ for each soil gives the larger $\Sigma$ value than $\Sigma_{min}$. This result of the $k$ value choice relates to the soils under consideration with $U_{lph}=0$ (Table 2). For soils with $U_{lph}>0$ the choice of $k>0$ would be preferable. The ranges obtained for $X_{mz}$ and $K$ (Table 3) are quite reasonable from the viewpoint of usually observed values. Figures 5-8 show the found reference shrinkage curves, $Y_r(W)$ of the four soils. The particular values of $x_m$, $X_{mz}$ and $K$ relating to the narrow ranges in Table 3 are indicated in the figure captions ($k=0$).

7. Using $\Sigma_{min}$ one can estimate the standard deviation, $\sigma_{Yr}$ (Hamilton, 1964) of the experimental specific soil volume $Y_{re}$ values (Figs.1-4) as $\sigma_{Yr}=(\Sigma_{min}/(N-1))^{1/2}$ where $N$ being the number of experimental points (see $\sigma_{Yr}$ in the captions of Figs.5-8). It should be noted that $\sigma_{Yr} \cong \delta Y_{re}$ where $\delta Y_{re}$ is the direct estimate of the standard error by the spread of the experimental points at a given water content in Figs.5-8.

Table 4 shows the values of all the intermediate soil characteristics that participate in the $K$ and $k$ calculation according to the general algorithm (see the Appendix) at $X_{mz}$ and $x_m$ indicated in captions of Figs.5-8. It is worth reiterating that with available texture and structure data ($x_m, X_{mz}$), the $K$ and $k$ soil property estimation and then the reference shrinkage curve estimation could be conducted simpler (see the Appendix) and independently of the experimental data in Figs.5-8 (white circles) at $0<W<W_h$.

*11.2. Estimates of the critical sample size ($h^*$)*

First, using Eq.(A6) we find the mean soil-solids size, $x_n$ of the soils. The $x_n$ size approximately coincides with the minimum aggregate size in the oven-dried state ($X_{\text{minz}}$) and at maximum swelling ($X_{\text{min}}$) (Chertkov, 2008b). The estimated $X_{\text{min}}$ values are indicated in Table 5. Then, the values of $x_n \cong X_{\text{min}}$ (Table 5), $X_{\text{mz}}$ (Table 5), $u_z$ and $u_h$ (Table 4) for the soils determine their maximum aggregate size at maximum swelling, $X_m$ as (Chertkov, 2008b)

$$X_m = x_n + (X_{\text{mz}} - x_n)(u_h/u_z)^{1/3} \ . \tag{42}$$

The $X_{\text{min}}$ and $X_m$ values (Table 5) permit the calculation of the mean distances, $l_{\text{min}}$ and $l_m$ between the smallest and largest aggregates, respectively, in the soils (Eqs.(23)-(24); Table 5). The estimated critical sample size, $h^*$ (Eqs.(25)-(26a)) is also given in Table 5 and used in the following section. The $h^*$ estimates for real soils from Crescimanno and Provenzano (1999) confirm the estimate values from Table 1.

*11.3. Analysis of large-sample data*

The specific volumes, $Y_s(W)$ and $Y_r(W)$ (Eq.(33)) are linked as

$$Y_s(W) = Y_r(W) + U_{\text{cr s}}(W) - U_s \ . \tag{43}$$

Combining $Y_r(W) = U(W)/K + U_i + U_s$ (Chertkov, 2007a) and Eq.(29) one can also express the specific crack volume of the sample, $U_{\text{cr s}}(W)$ through the sample crack factor $q_s(h/h^*)$, $Y_r(W)$, and the constant (for the soil) values, $U_s$, $U_i$, $U_h$ as

$$U_{\text{cr s}}(W) = -q_s Y_r(W) + q_s(U_h/K + U_s + U_i) + U_s = -q_s Y_r(W) + q_s Y_{rh} + U_s \ . \tag{44}$$

The *first aim* of the section is to validate *assumption* 1 (Eq.(12) and the end of section 3), that is, to check the feasibility of the above $Y_s$ presentation (Eqs.(43) and (44)) as a function of $W$ through the constant $q_s$ factor at $h/h^* > 1$. We are based on the data on four soils from Crescimanno and Provenzano (1999), obtained using the large samples (cores of size $h$=11.5cm at $h^*$ from Table 5) and presented in Figs.5-8 by black circles. The $U_h$, $U_i$ (Table 4), and $K$ values (see the captions to Figs.5-8) as well as the $Y_r(W)$ dependence (the lower solid curve in Figs.5-8) for the soils were found above (for all the soils $U_s$=0). To present the experimental data for each soil (black circles in Figs.5-8) using the $Y_s(W)$ dependence from Eqs.(43)-(44), we took advantage of the least-square criterion and found the best-fit $q_s$ values (Table 6). Figures 5-8 show by the upper solid lines the found shrinkage curves, $Y_s(W)$ of large samples. The goodness of fit, $r^2_{Ys}$ varies between 0.77 and 0.98 for the four soils, and the estimate $\sigma_{Ys}$ of the standard deviation of the experimental values of $Y_s$ (black circles in Figs.5-8) varies between 0.0138 and 0.0170 dm$^3$kg$^{-1}$ (for the exact $r^2_{Ys}$ and $\sigma_{Ys}$ values see the captions of Figs.5-8). The slopes, $S$ of the shrinkage curves, $Y_s(W)$ in the basic shrinkage range (i.e., at $W_n \leq W \leq W_s$ in Figs.5-8), were found at $k$=0 and $q_s$ from Table 6 (see Eq.(17)) and are also given in Table 6.

The following specification should be noted. The data (black circles) in Figs.6-8 contain points that lie on the saturation line at $W > W_h$. The "shrinkage" along the saturation line corresponds to possible closing of the water filled capillary cracks between aggregates at drying, but not to inter-aggregate macro-cracking (Chertkov and Ravina, 2001). For this reason in the above fitting the only experimental points (black circles in Figs.5-8) that were used were those that lie at $W \leq W_h$.

Note that the above standard deviations, $\sigma_{Ys}$ coincide by the order of magnitude with the immediate estimates of experimental errors, $\delta Y_{se}$ from the spread of the experimental points (black circles) in Figs.5-8. Therefore, each fitted curve lies within the limits of the experimental errors of the $Y_{se}$ values. That is, the found shrinkage curves with cracks, $Y_s(W)$ are in agreement with the experimental data not only from the viewpoint of the fitting criterion (connected with the $r^2{}_{Ys}$ values), but also from the viewpoint of the standard physical criterion (the predicted curve is within the limits of experimental errors). This sufficiently good agreement speaking in favor of the feasibility of the crack factor concept and its independence from the water content (assumption 1) takes place despite the quite large spread of the experimental points (black circles in Fig.5-8). In addition to the mean $q_s$ values (Table 6) we estimated their standard errors, both positive, $\delta q_{s+}$ and negative, $\delta q_{s-}$ (Table 6). To this end we estimated the best-fit $q_s$ values for the experimental $Y_s$ points (black circles in Figs.5-8) after their displacement up and down by the above found standard deviation, $\sigma_{Ys}$. The $\delta q_{s+}$ and $\delta q_{s-}$ values will be used below. Finally, note that according to Eq.(43) (at $U_s$=0) the difference between the $Y_s(W,h/h^*)$ and $Y_r(W)$ curves in Figs.5-8 gives the specific crack volume $U_{cr\,s}(W,h/h^*)$ for the large soil samples (cores)

The *second aim* of this section is to validate *assumption* 2 (Eq.(26a)) and *assumption* 3 (Eqs.(27) with the indicated $b_1$, $b_2$, and $\delta$ values), that is, to check (using the available limited data) the presentation feasibility of the sample crack factor, $q_s$ as a function of the relative sample size, $h/h^*$ (Eqs.(27a-d)). We used the four points in coordinates ($h/h^*$, $q_s$) (Table 6; Fig.9) estimated above for the four soils at $h$=11.5cm, and the standard-error estimates, $\delta q_{s+}$ and $\delta q_{s-}$ of $q_s$ (Table 6 and vertical bars in Fig.9). The $q_s(h/h^*)$ dependence (Eq.(27)) at the theoretical values, $b_1$=0.15, $b_2$=1, $\delta$=1.5 is shown in Fig.9 by the solid line. One can see that the deflection between the theoretical $q_s(h/h^*)$ curve and ($h/h^*$, $q_s$) points found for the four soils is within the limits of the two standard errors. Hence, from the viewpoint of the standard physical criterion the theoretical $q_s(h/h^*)$ curve does not contradict the available data from Crescimanno and Provenzano (1999). Evaluating the results of the $q_s(h/h^*)$ curve (and correspondingly assumptions 2 and 3) validation one should take into account the following error sources: (i) uncertainties in $h$ ($\delta h$=3cm=11.5-8.5 cm) and $h^*$ (the $10^3$ coefficient in Eq.(26a)); (ii) uncertainties in $q_s$ (the vertical bars in Fig.9); (iii) the small number of soils and corresponding small number of points in Fig.9. Uncertainties in $h$ can be essentially decreased using cores of similar height and diameter. Uncertainties in $h^*$ can be essentially decreased using the immediately measured characteristic $X_m$ and $x_m$ sizes. Uncertainties in $q_s$ can be essentially decreased using the more accurate data on the shrinkage of large samples. The number of soils used in the estimation can be obviously increased with time. However, in spite of these error sources, as said, the above comparison (Fig.9) is quite satisfactory. The accumulation of relevant data on many soils will enable one to additionally specify $b_1$ in Eq.(27b) ($b_2$ and $\delta$ depend on $b_1$) and the coefficient (close to $10^3$) in Eq.(26a) as universal parameters. Thus, one can preliminarily state that the above results show evidence in favor of assumptions 2 and 3 for the sample case, although additional checking and specification are desirable. According to Eq.(28) we can use the corresponding universal theoretical dependence, $q_l(h/h^*)$ for the layer case with the same $b_1$, $b_2$, and $\delta$ (see dashed line in Fig.4). The illustrative examples of the layer shrinkage curves, $Y_l(W)$ predicted for two different layer thicknesses $h<h^*$ and $h>h^*$ (for $h^*$ see Table 5) are shown in Figs.6 and 7 for Delia 2a and Delia 4 soils.

## 12. Results and discussion

The theoretical results are as follows. (i) New concepts of the lacunar factor ($k$), crack factor ($q$), and critical sample size ($h^*$) were introduced. (ii) The expression for $h^*$ through parameters of aggregate-size distribution was derived. (iii) The expressions $q_s(h/h^*)$ and $q_l(h/h^*)$ were derived for the crack factors in the case of sample ($q_s$) and layer ($q_l$) shrinkage ($h$ being the initial sample size or layer thickness). (iv) The expression for the lacunar factor $k$ through clay content (Chertkov, 2010b) was generalized to the case of shrinkage with cracking. (v) Based on the reference shrinkage curve of the soil, $Y_r(W)$ as well as the expressions of $q_s$ and $q_l$, there were found the expressions for the shrinkage curves of the sample of a given size, $Y_s(W, h/h^*)$ and layer of a given thickness, $Y_l(W, h/h^*)$, accounting for the crack volume for these cases, $U_{cr\,s}(W, h/h^*)$ and $U_{cr\,l}(W, h/h^*)$.

The results of the new concepts validation are as follows

(i) *The estimates of texture and structure characteristics*. Besides the direct measurements of $x_m$ and $X_{mz}$ characteristics, they can be sufficiently accurately estimated by the method stated in section 11.1, simultaneously with the estimates of the $K$ ratio and lacunar factor, $k$, even at an essential spread of the experimental shrinkage curve points. The physically reasonable $x_m$, $X_{mz}$, $K$, and $k$ values in Table 3 and the above use of these values evidence in favor of the statement.

(ii) *The prediction of the critical sample size through soil texture and structure*. The estimation of the critical sample size (Section 11.2) for the soils from Crescimanno and Provenzano (1999), showed the physically reasonable estimates of $h^*$ (Table 5) that are close to the theoretical ones (Table 1). The estimates of the $l_{min}/X_{min}$ and $l_m/X_m$ ratios for the real soils (Table 5) also confirm the theoretical estimates (Table 1).

(iii) *Presentation of the shrinkage curve and crack volume through the crack factor*. Section 11.3 shows evidence in favor of the feasibility of the shrinkage curve (Eq.(43)) and specific crack volume (Eq.(44)) presentation through the $q$ factor that does not depend on water content for both samples and layers.

(iv) *Presentation of the crack factor through the relative sample size/layer thickness*. Section 11.3 also shows evidences in favor of the feasibility of the sample crack factor presentation (Eq.(27)) through the $h/h^*$ ratio, and, taking into account the relation between $q_s$ and $q_l$, the same relates to the layer crack factor (Eq.(28)).

Thus, the approach taking into account the sample size/layer thickness as well as soil texture and structure at shrinkage and cracking prediction, permits one to introduce corresponding soil features as physical ones. The $h$ size is taken into account immediately through the $h/h^*$ ratio as an argument of the crack factor, $q$. The soil texture and structure influence the shrinkage curve and crack volume at drying through: (i) the $K$ ratio dependence on $x_n \cong X_{min}$, $X_m$, and aggregate-size distribution (Chertkov, 2008b), (ii) the lacunar $k$ factor dependence on the $c/c_*$ ratio (in turn, $c_*$ depends on $p$, $v_s$, and $v_z$) (Chertkov, 2010b), and (iii) the critical size $h^*$ dependence on $x_n \cong X_{min}$, $X_m$, and aggregate-size distribution (Section 6).

## 13. Conclusion

In connection with the physical description of shrinkage and cracking of soil samples (layers) of different size, texture, and structure, the concepts of the critical sample size, lacunar factor, and crack factor were introduced. The results of the present work give a primary experimental confirmation of the concepts' feasibility. This confirmation is reached by the analysis of the relevant data (Crescimanno and Provenzano, 1999) on the shrinkage curves of four soils, using "small" and "large" samples. Thereby, the recently introduced model of the intra-aggregate soil structure (Chertkov, 2007a, 2007c, 2008a) that underlies the approach under consideration, is also additionally confirmed. The limitedness of the experimental data that are relevant to the aims of this work is evident. For this reason additional experimental checking is desirable. Nonetheless, the results of the above analysis are promising. Implying the hydrological applications of the obtained results, it is worth noting the necessity of the additional generalization of the above new concepts to the swelling case (beyond the scope of this work), to physically and quantitatively model non-total crack closing at the following swelling and the maximum-swelling-volume decrease after a shrink-swell cycle at the seasonal shrink-swell rotation.

**Appendix: General algorithm of the $K$ and $k$ estimation**

The necessary soil features to find $K$ and $k$ are characterized by parameters from Table 2 as well as $X_{mz}$ and $x_m$. First, one finds the $K$ ratio.

(1) The input parameters (Table 2) successively give (Chertkov, 2007a, 2007c) (for value meaning see Notation):

$$u_s = 1/(1+2\rho_s W_h) \,, \tag{A1}$$

$$u_{lph} = U_{lph} u_s \rho_s \,, \tag{A2}$$

$$u_h = 0.5(1+u_s) + u_{lph} \,, \tag{A3}$$

$$U_h = u_h/(u_s \rho_s) \,, \tag{A4}$$

$$U_s = P_z Y_{rz} \,. \tag{A5}$$

(2) Using the primary textural data (Table 2) and taking clay particles in the size range, $x \leq 0.002$ mm, silt grains in the size range, $0.002 < x \leq 0.050$ mm, and sand grains in the size range, $0.050 < x \leq x_m$, one estimates the mean soil solids size, $x_n$ (mm) to be

$$x_n = 0.001c + 0.026 s_1 + (0.025 + x_m/2) s_2 \,. \tag{A6}$$

(3) The found parameters, $u_s$ (Eq.(A1)), $u_h$ (Eq.(A3)), $U_h$ (Eq.(A4)), $U_s$ (Eq.(A5)), together with parameters from Table 2 ($Y_{rz}$, $\rho_s$, and $P_z$) as well as $x_n$ (Eq.(A6)) and $X_{mz}$, enable one to solve the non-linear equation for $K$ (Chertkov, 2008b) with any accuracy.

Then one finds the $k$ factor.

(4) The parameters from Table 2 ($c$, $Y_{rz}$, $\rho_s$), the found $u_s$ (Eq.(A1)), $U_h$ (Eq.(A4)), $U_s$ (Eq.(A5)) values as well as the found $K$ value successively give (Chertkov, 2007a, 2007c) parameters that enter the equation eventually determining $k$ (see below; for value meaning see Notation):

$$u_S = u_s(1-c) \,, \tag{A7}$$

$$v_s = (u_s - u_S)/(1 - u_S) \,, \tag{A8}$$

$$v_h = 0.5(1+v_s) \,, \tag{A9}$$

$$U_i = U_h(1 - 1/K) \,, \tag{A10}$$

$$u_z = (Y_{rz} - U_i - U_s) K u_s \rho_s \,. \tag{A11}$$

(5) Porosity $p$ of contributive silt and sand grains in the state of *imagined* contact (that enter the equation eventually determining $k$) is found as the solution of Eq.(40) with taking into account Eqs.(36) and (39) as well as the $f_{silt}$ estimate as

$$f_{silt} = s_1/(s_1 + s_2) \,. \tag{A12}$$

(6) $k$ is found from Eq.(19) (accounting for $k(c/c_*)=0$ at $1 \leq c/c_* < 1/c_*$) and $c_*$ from Eq.(18). On the other hand, the relative clay volume in the oven-dried state, $v_z$ can be written as (Chertkov, 2007c)

$$v_z = (u_z - u_S - u_{lph})/(1-u_S) - k(v_h - v_z) \ . \tag{A13}$$

Replacing here $k$ with $k(c/c_*(v_z))$ from Eqs.(19) and (18) and denoting after that the right part of Eq.(A13) by $v_z'(v_z)$, we come to the non-linear equation relative to $v_z$ as

$$v_z = v_z'(v_z) \ . \tag{A14}$$

This equation is solved numerically (at found $u_z$, $u_s$, $u_{lph}$, $u_S$, $v_h$, $v_s$, and $p$). Then the found $v_z$ value gives $c_*(v_z)$ (Eq.(18)) and $k$ (Eq.(19)). The numerical analysis shows that the solution of Eq.(A14) at given primary and input parameters (see Table 2) exists not for any possible $X_{mz}$ and $x_m$ values (i.e., not for any possible structure and texture). The analysis and its results are considered in Section 11.1. Finally, it is worth emphasizing that the above algorithm is relevant to the general situation with $P_z \neq 0$ and $U_{lph} \neq 0$, unlike the particular situation we deal with in the data analysis (see Table 2) in the force of the limitations of relevant available data.

**Notation**

$b_1$, $b_2$    coefficients in Eqs.(27) and (28) (dimensionless)
$c$, $c_*$    clay content and critical clay content (dimensionless)
$F(X)$    cumulative aggregate-size distribution (dimensionless)
$f(x)$    silt and sand grain-size distribution (dimensionless)
$f_{silt}$    $f$ value at $\eta = \eta_{silt}$ (Eq.(40)) (dimensionless)
$h$    initial sample size or layer thickness at maximum swelling (cm)
$h^*_o$    rough approximation of the critical sample size at maximum swelling, $h^*$ (cm)
$I_o(\eta)$    function from Eq.(36) (dimensionless)
$K$    ratio of aggregate solid mass to that of intra-aggregate matrix (dimensionless)
$k$    lacunar factor (dimensionless)
$l$    mean distance between the aggregates of size $X$ ($l_m$ at $X_m$, $l_{min}$ at $X_{min}$) (mm)
$P_h$    inter-aggregate porosity at maximum swelling (dimensionless)
$p$    porosity of contacting silt and sand grains (dimensionless)
$q$    crack factor of sample ($q_s$) or layer ($q_l$) (dimensionless)
$r^2_{Yr}$    goodness of fit of $Y_r(W)$ at $X_{mz}$ variation (dimensionless)
$r^2_{Ys}$    goodness of fit of $Y_s(W)$ at $q$ variation (dimensionless)
$S$, $S_r$    shrinkage or reference shrinkage curve slope in the basic range (dm$^3$ kg$^{-1}$)
$s_1$, $s_2$    silt and sand content (dimensionless)
$U(w)$    specific volume of intra-aggregate matrix per unit mass of the oven-dried matrix itself (dm$^3$ kg$^{-1}$)
$U_h$, $U_z$    $U$ value at maximum swelling and shrinkage limit (Fig.2) (dm$^3$ kg$^{-1}$)
$U_a(w')$    specific volume of aggregates (per unit mass of the oven-dried soil) (dm$^3$ kg$^{-1}$)
$U_{ah}$, $U_{az}$    $U_a$ value at maximum swelling and shrinkage limit (Fig.2) (dm$^3$ kg$^{-1}$)
$U_{cp}(w)$    similar to $U'_{cp}(w')$, but related to the unit mass of the oven-dried intra-aggregate matrix itself (dm$^3$ kg$^{-1}$)
$U_{cr}(w')$    specific volume of cracks (per unit mass of the oven-dried soil) (dm$^3$ kg$^{-1}$)
$U_{cr\,s}$, $U_{cr\,l}$    specific crack volume of sample or layer (dm$^3$ kg$^{-1}$)
$U_{cs}$    similar to $U'_{cs}$, but related to the unit mass of the oven-dried intra-aggregate matrix itself ($U_{cs} \equiv 1/\rho_s$) (dm$^3$ kg$^{-1}$)

$U_i$     specific volume of interface layer per unit mass of the oven-dried soil (dm$^3$ kg$^{-1}$)

$U_{lp}(w)$ similar to $U'_{lp}(w')$, but related to unit mass of the oven-dried intra-aggregate matrix itself (dm$^3$ kg$^{-1}$)

$U_{lph}$     specific lacunar pore volume at maximum swelling (dm$^3$ kg$^{-1}$)

$U_{lpz}$     specific lacunar pore volume at shrinkage limit (dm$^3$ kg$^{-1}$)

$U_s$ specific volume of structural pores per unit mass of the oven-dried soil (dm$^3$ kg$^{-1}$)

$U'(w')$ specific volume of intra-aggregate matrix (per unit mass of the oven-dried soil) (dm$^3$ kg$^{-1}$)

$U'_{cp}(w')$     specific volume of clay pores of intra-aggregate matrix (per unit mass of the oven-dried soil) (dm$^3$ kg$^{-1}$)

$U'_{cs}$     specific volume of solids of intra-aggregate matrix (per unit mass of the oven-dried soil) (dm$^3$ kg$^{-1}$)

$U'_{lp}(w')$     specific volume of lacunar pores of intra-aggregate matrix (per unit mass of the oven-dried soil) (dm$^3$ kg$^{-1}$)

$u_h, u_z$ relative volume of an intra-aggregate matrix at maximum swelling and shrinkage limit (dimensionless)

$u_{lph}$ relative volume of lacunar pores in the intra-aggregate matrix at maximum swelling (dimensionless)

$u_S, u_s$ relative volume of non-clay solids and all solids of an intra-aggregate matrix (dimensionless)

$v_h, v_z$ relative volume of clay at maximum swelling and shrinkage limit (dimensionless)

$v_s$     relative volume of clay solids (dimensionless)

$W$     gravimetric water content of soil (kg kg$^{-1}$)

$W_m, W_h, W_s, W_n, W_z$     $W$ value at filling of the structural pores, maximum swelling, end point of structural shrinkage, end point of basic shrinkage, and shrinkage limit (kg kg$^{-1}$)

$W_h^*, W_m^*$     water content corresponding to the $U_a=U_{ah}$ and $Y=Y_h$ value on the true saturation line in Fig.2 (kg kg$^{-1}$)

$w$     the water content of the intra-aggregate matrix per unit mass of the oven-dried matrix itself (kg kg$^{-1}$)

$w_n, w_z$     $w$ value at air-entry point and at shrinkage limit (Fig.2) (kg kg$^{-1}$)

$w'$     water content of intra-aggregate matrix (per unit mass of the oven-dried soil) (kg kg$^{-1}$)

$w'_h$     $w'$ value at maximum swelling (kg kg$^{-1}$)

$X$     aggregate size (mm)

$X_m, X_{min}$     maximum and minimum aggregate size at maximum swelling (mm)

$x_m, x_{min}$     maximum and minimum grain size (μm or mm)

$x_n$     mean soil-solids size (μm or mm)

$x_{m\,max}, x_{m\,min}$     upper and lower border of the $x_m$ values in Table 3 (mm)

$Y$     specific volume of the soil with cracks (dm$^3$ kg$^{-1}$)

$Y_h$     specific soil volume at maximum swelling (dm$^3$ kg$^{-1}$)

$Y_l, Y_s$     specific soil volume in case of layer and sample (dm$^3$ kg$^{-1}$)

$Y_r(w')$     reference shrinkage curve (dm$^3$ kg$^{-1}$)

$Y_{re}$     experimental values of the reference shrinkage curve, $Y_r(W)$ (dm$^3$ kg$^{-1}$)

$Y_{rz}$     oven-dried specific volume of soil without cracks (dm$^3$ kg$^{-1}$)

$Y_{se}$     experimental values of the sample shrinkage curve, $Y_s(W)$ (dm$^3$ kg$^{-1}$)

$\Delta X$     range of aggregate sizes (mm)

$\delta$     constant in Eqs.(27) and (28) (dimensionless)

$\delta q_{s+}, \delta q_{s-}$     positive and negative standard deviations of $q_s$ values (dimensionless)

δ$Y_{re}$, δ$Y_{se}$   experimental errors of the $Y_{re}$ and $Y_{se}$ values (dm$^3$ kg$^{-1}$)
η   parameter from Eqs.(36) and (38) (dimensionless)
η$_{silt}$   parameter from Eq.(39) (dimensionless)
ρ$_s$, ρ$_w$   mean solid density and density of water (kg dm$^{-1}$)
σ$_{Yr}$, σ$_{Ys}$   standard deviation of $Y_{re}$ and $Y_{se}$ values (dm$^3$ kg$^{-1}$)
ω($w'$)   water content of interface layer (per unit mass of the oven-dried soil) (kg kg$^{-1}$)

**References**

Bear, J., Bachmat, Y., 1990. Introduction to Modeling of Transport Phenomena in Porous Media. Kluwer Academic Publishers, London.
Boivin, P., Garnier, P., Vauclin, M., 2006. Modeling the soil shrinkage and water retention curves with the same equations. Soil Sci Soc. Am J. 70, 1082-1093.
Braudeau, E., Mohtar, R.H., 2004. Water potential in nonrigid unsaturated soil-water medium. Water Resour. Res. 40, W05108, doi: 10.1029/2004WR003119.
Braudeau, E., Frangi, J-P., Mohtar, R.H., 2004. Characterizing nonrigid aggregated soil-water medium using its shrinkage curve. Soil Sci. Soc. Am. J. 68, 359-370.
Braudeau, E., Sene, M., Mohtar, R.H., 2005. Hydrostructural characteristics of two African tropical soils. Europ. J. Soil Sci. 56, 375-388.
Chertkov, V.Y., 2000a. Using surface crack spacing to predict crack network geometry in swelling soils. Soil Sci. Soc. Am. J. 64, 1918-1921.
Chertkov, V.Y., 2000b. Modeling the pore structure and shrinkage curve of soil clay matrix. Geoderma 95, 215-46.
Chertkov, V.Y., 2003. Modelling the shrinkage curve of soil clay pastes. Geoderma 112, 71-95.
Chertkov, V.Y., 2005a. The shrinkage geometry factor of a soil layer. Soil Sci. Soc. Am. J. 69, 1671-1683.
Chertkov, V.Y., 2005b. Intersecting-surfaces approach to soil structure. Int Agrophysics 19, 109-118.
Chertkov, V.Y., 2007a. The reference shrinkage curve at higher than critical soil clay content. Soil Sci. Soc. Am. J. 71, 641-55.
Chertkov, V.Y., 2007b. The reference shrinkage curve of clay soil. Theor. and Appl. Fracture Mechanics 48, 50-67.
Chertkov, V.Y., 2007c. The soil reference shrinkage curve. Open Hydrology Journal 1, 1-18.
Chertkov, V.Y., 2008a. The physical effects of an intra-aggregate structure on soil shrinkage. Geoderma 146, 147-56.
Chertkov, V.Y., 2008b. Estimating the aggregate/intraaggregate mass ratio of a shrinking soil. Open Hydrology Journal 2, 7-14.
Chertkov, V.Y., 2010a. Two-factor model of soil suction from capillarity, shrinkage, adsorbed film, and intra-aggregate structure. Open Hydrology Journal 4, 44-64.
Chertkov, V.Y., 2010b. The soil lacunar factor and reference shrinkage. International Agrophysics 24, 227-235.
Chertkov, V.Y., Ravina, I., 1998. Modeling the crack network of swelling clay soils. Soil Sci. Soc. Am. J. 62, 1162-1171.
Chertkov V.Y., Ravina, I., 1999. Tortuosity of crack networks in swelling clay soils. Soil Sci. Soc. Am. J. 63, 1523-1530.
Cornelis, W.M., Corluy, J., Medina, H., Diaz, J., Hartmann, R., Van Meirvenne, M., Ruiz, M.E., 2006. Measuring and modeling the soil shrinkage characteristic curve. Geoderma 137, 179-191.

Crescimanno, G., Provenzano, G., 1999. Soil shrinkage characteristic curve in clay soils: Measurement and prediction. Soil Sci. Soc. Am. J. 63, 25-32.

Dasog, G.S., Acton, D.F., Mermut, A.R., De Jong, E., 1988. Shrink-swell potential and cracking in clay soils of Saskatchewan. Can. J. Soil Sci. 68, 251-260.

Fiès, J.C., Bruand, A., 1998. Particle packing and organization of the textural porosity in clay-silt-sand mixtures. Eur. J. Soil Sci. 49, 557-567.

Gdoutos, E.E., 1993. Fracture Mechanics. Kluwer Academic Publishers, London.

Hamilton, W.C., 1964. Statistics in physical science. Ronald Press: New York.

McGarry, D., Daniels, I.J., 1987. Shrinkage curve indices to quantify cultivation effects of soil structure of a Vertisol. Soil Sci. Soc. Am. J. 51, 1575-1580.

Olsen, P.A., Haugen, L.E., 1998. A new model of the shrinkage characteristic applied to some Norwegian soils. Geoderma 83, 67-81.

Peng, X., Horn, R., 2005. Modeling soil shrinkage curve across a wide range of soil types. Soil Sci. Soc. Am. J. 69, 584-592.

Tariq, A.-R., Durnford, D.S., 1993. Analytical volume change model for swelling clay soils. Soil Sci. Soc. Am. J. 57, 1183-1187.

Yaalon, D.H., Kalmar, D., 1984. Extent and dynamics of cracking in a heavy clay soil with xeric moisture regime. In J.Bouma and P.A.C. Raats (Eds.) Proc. ISSS Symp. Water and solute movement in heavy clay soils, Wageningen, the Netherlands. 27-31 Aug. 1984. ILRI, Wageningen, the Netherlands, pp.45-48.

Yule, D.F., Ritchie, J.T., 1980a. Soil shrinkage relationships of Texas vertisols: I. Small cores. Soil Sci. Soc. Am. J. 44, 1285-1291.

Yule, D.F., Ritchie, J.T., 1980b. Soil shrinkage relationships of Texas vertisols: I. Large cores. Soil Sci. Soc. Am. J. 44, 1291-1295.

Zein el Abedine, A., Robinson, G.H., 1971. A study on cracking in some Vertisols of the Sudan. Geoderma 5, 229-241.

**Figure captions**

**Fig.1.** The schematic illustration of the accepted soil structure (Chertkov, 2007a, 2007c). Shown are (1) an assembly of many soil aggregates and inter-aggregate pores contributing to the specific soil volume, $Y$; (2) an aggregate, as a whole, contributing to the specific volume $U_a=U_i+U'$; (3) an aggregate indicated with two parts: (3a) interface layer contributing to the specific volume $U_i$ and (3b) intra-aggregate matrix contributing to the specific volumes $U$ and $U'=U/K$; (4) an aggregate indicated with the intra-aggregate structure: (4a) clay, (4b) silt and sand grains, and (4c) lacunar pores; and (5) an inter-aggregate pore leading, at shrinkage, to inter-aggregate crack contributing to the specific volume $U_{cr}$. $U$ is the specific volume of the intra-aggregate matrix (per unit mass of the oven-dried matrix itself). $U'$ is the specific volume of the intra-aggregate matrix (per unit mass of the oven-dried soil). $U_i$ is the specific volume of the interface layer (per unit mass of the oven-dried soil). $U_{cr}$ is the specific volume of cracks (per unit mass of the oven-dried soil). $U_a$ is the specific volume of aggregates (per unit mass of the oven-dried soil). $K$ is the aggregate/intra-aggregate mass ratio.

**Fig.2.** The scheme of the step-by-step transition from the specific volume of the intra-aggregate matrix ($U(W)$) to the specific volume of the cracked soil ($Y(W)$): $U \rightarrow U' \rightarrow U_a \rightarrow Y_r \rightarrow Y$ ($1 \rightarrow 2 \rightarrow 3 \rightarrow 4 \rightarrow 5$). $U$ and $U'$ are the specific volumes of the intra-aggregate matrix normalized to the matrix mass and to the mass of the soil as a whole, respectively. $S_r=(1-k)/\rho_w$ is the slope of the reference shrinkage curve, $Y_r(W)$ in the basic shrinkage range. $S=(1-q)(1-k)/\rho_w$ is the slope of $Y(W)$ in the basic shrinkage range. $U_s$ is the specific volume of inter-aggregate pores at maximum swelling. $U_i$ and $U_{cr}$ as in Fig.1. $U_{lph}$ is the specific volume of lacunar pores at maximum swelling. Dashed and dash-dot inclined straight lines are the true and pseudo saturation lines, respectively for both the $W$ and $w$ axes. $W_h^*$ and $W_m^*$ are the water contents corresponding to $U_a=U_{ah}$ and $Y=Y_h$ values, respectively, on the true saturation line.

**Fig.3.** The solid curve simultaneously illustrates the qualitative view and the analytical dependence of $k(c/c_*)$.

**Fig.4.** The solid and dashed curves simultaneously illustrate the qualitative view and the analytical dependence of $q_s(h/h^*)$ and $q_l(h/h^*)$ for samples and layers, respectively, at theoretically estimated $b_1 \cong 0.15$ and corresponding $b_2 \cong 1$ and $\delta \cong 1.5$.

**Fig.5.** White circles, black circles, and thin solid inclined line are primary experimental data on small-sample shrinkage curve, large-sample shrinkage curve, and saturation line, respectively, for Delia 1a soil from Chescimanno and Provenzano (1999). The solid curved line, $Y_r(W)$ is the found reference shrinkage curve or shrinkage curve for small samples (clods) at $X_{mz}=4.5$mm and $x_m=0.075$mm. These $X_{mz}$ and $x_m$ lead to $k=0$ and $K=1.0363$, $r^2_{Yr}=0.9905$ and $\sigma_{Yr}=0.0129$ dm$^3$kg$^{-1}$. The solid curved line, $Y_s(W)$ is the found shrinkage curve for large samples (cores), including crack volume, at $q_s=0.223$, $r^2_{Ys}=0.9800$, $\sigma_{Ys}=0.0138$dm$^3$kg$^{-1}$. The black squares show the characteristic points on the shrinkage curves: $W_h$ is the maximum swelling point, $W_s$ is the end point of the structural shrinkage, $W_n$ is the end point of the basic shrinkage, $W_z$ is the shrinkage limit.

**Fig.6.** As in Fig.5, but for Delia 2a soil from Chescimanno and Provenzano (1999). $X_{mz}=3.65$mm, $x_m=0.160$mm, $k=0$, $K=1.2185$, $r^2_{Yr}=0.9840$, $\sigma_{Yr}=0.0089$ dm$^3$kg$^{-1}$, $q_s=0.494$, $r^2_{Ys}=0.8012$, $\sigma_{Ys}=0.0152$ dm$^3$kg$^{-1}$. In addition, the dashed curved lines show the predicted shrinkage curves for the "thin" layer, $Y_{l1}(W)$ at $h=3.1$cm ($h<h^*$) and the "thick" layer, $Y_{l2}(W)$ at $h=11.5$ cm ($h>h^*$).

**Fig.7.** As in Fig.6, but for Delia 4 soil from Chescimanno and Provenzano (1999). $X_{mz}$=5mm, $x_m$=0.160mm, $k$=0, $K$=1.1699, $r^2_{Yr}$=0.9463, $\sigma_{Yr}$=0.0144 dm$^3$kg$^{-1}$, $q_s$=0.280, $r^2_{Ys}$=0.8585, $\sigma_{Ys}$=0.0170 dm$^3$kg$^{-1}$.

**Fig.8.** As in Fig.5, but for Delia 6 soil from Chescimanno and Provenzano (1999). $X_{mz}$=3.60mm, $x_m$=0.130mm, $k$=0 and $K$=1.1746, $r^2_{Yr}$=0.9804, $\sigma_{Yr}$=0.0115 dm$^3$kg$^{-1}$, $q_s$=0.655, $r^2_{Ys}$=0.7705, $\sigma_{Ys}$=0.0151 dm$^3$kg$^{-1}$.

**Fig.9.** Black circles are the estimated points in coordinates of the relative sample size, $h/h^*$ and sample crack factor, $q_s$ (Table 6) for the four soils under consideration. The solid line shows the theoretical $q_s(h/h^*)$ dependence from Eq.(27) at $b_1 \cong 0.15$, and corresponding $b_2 \cong 1$ and $\delta \cong 1.5$ ($h/h^*=1+\delta$ is the "sewing" point). The vertical bars show the estimates of the standard deviations, $\delta q_{s+}$ and $\delta q_{s-}$ of $q_s$ (Table 6).

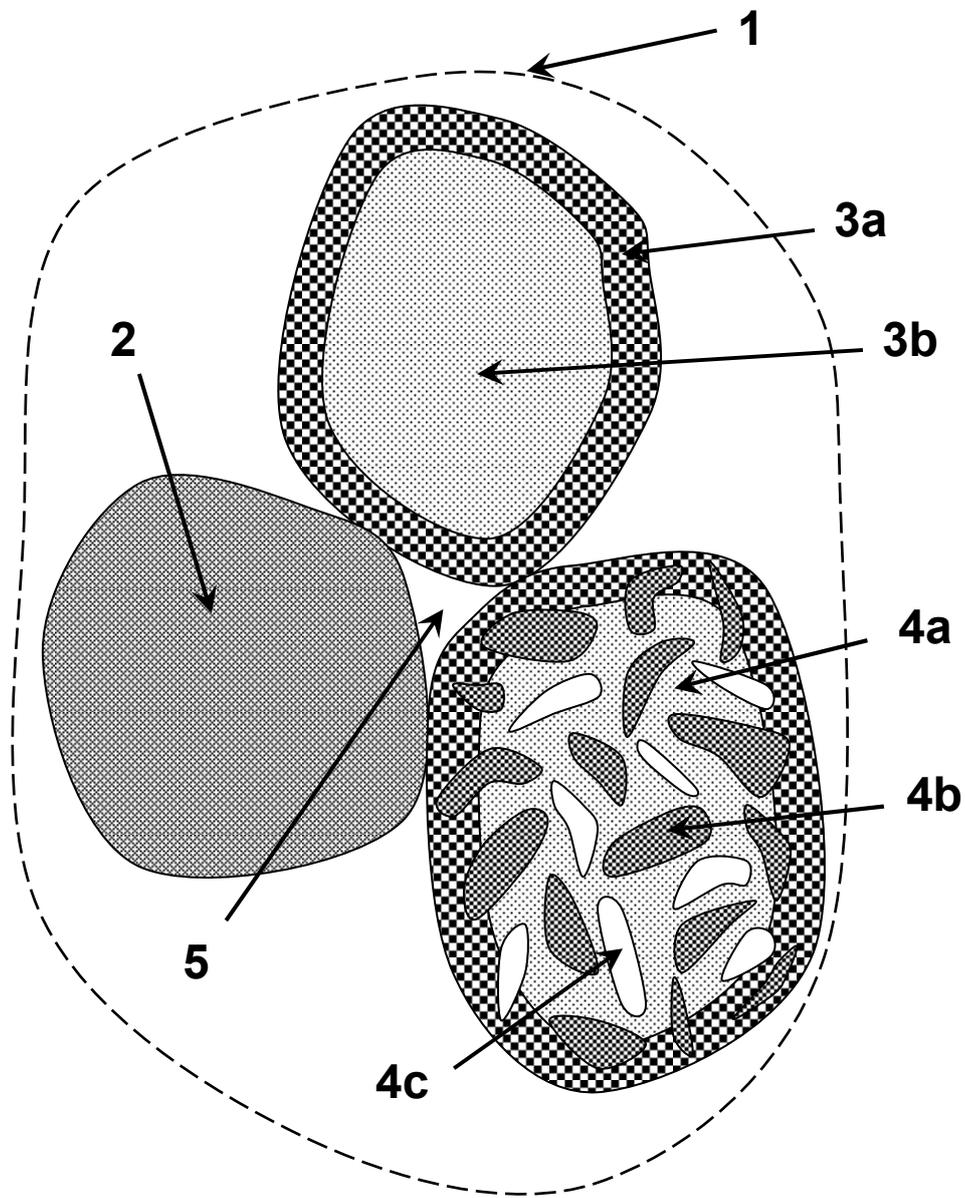

Fig.1

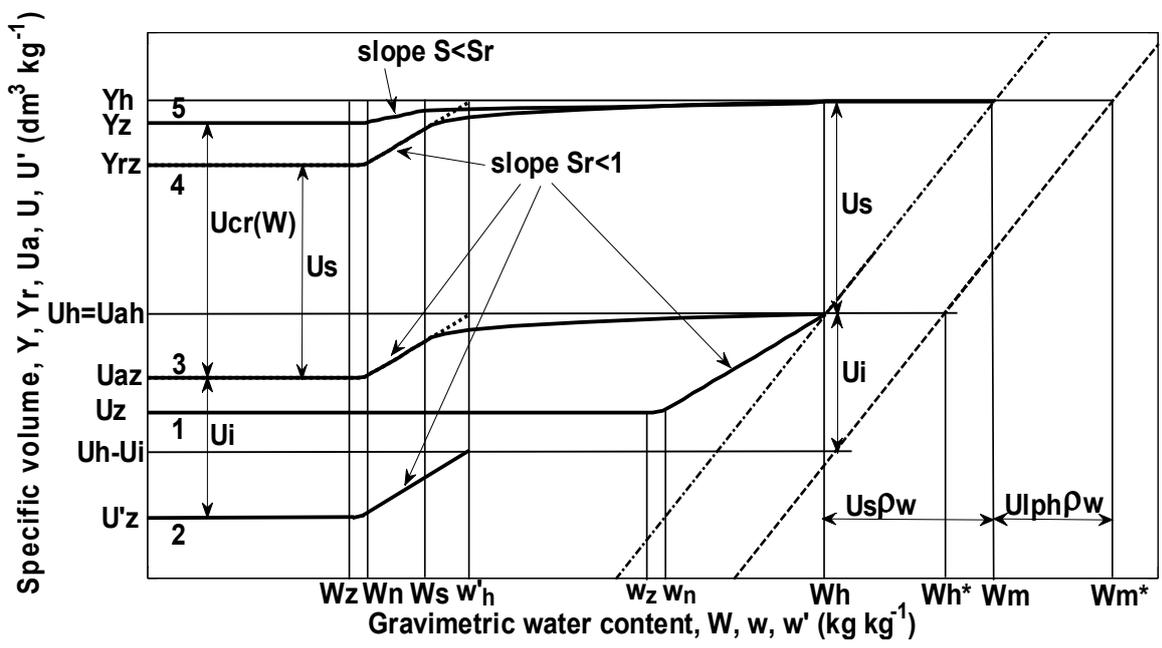

Fig.2

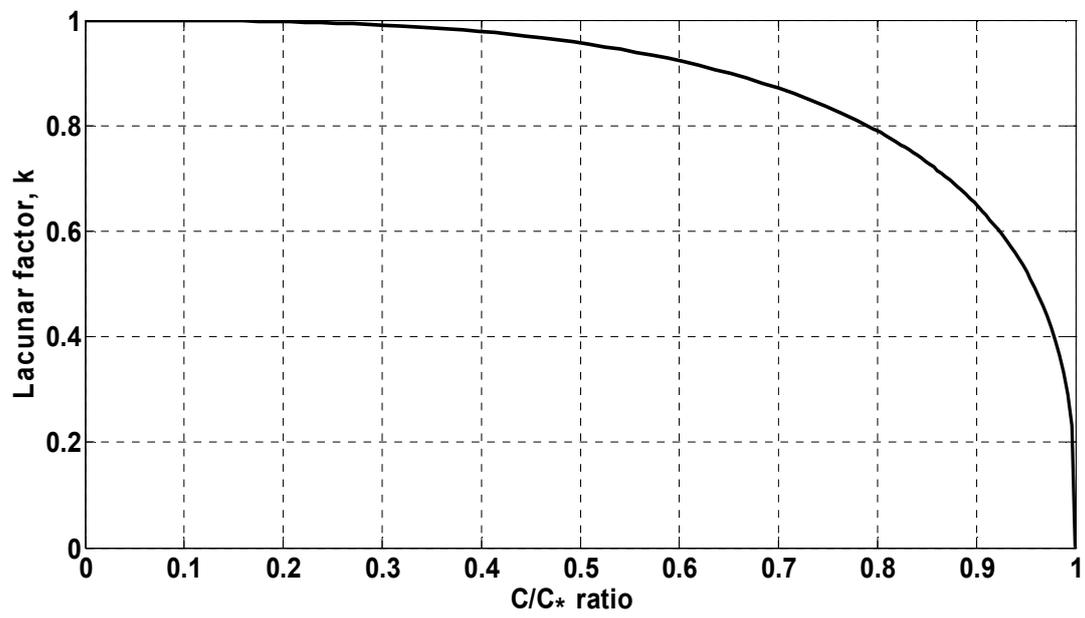

Fig.3

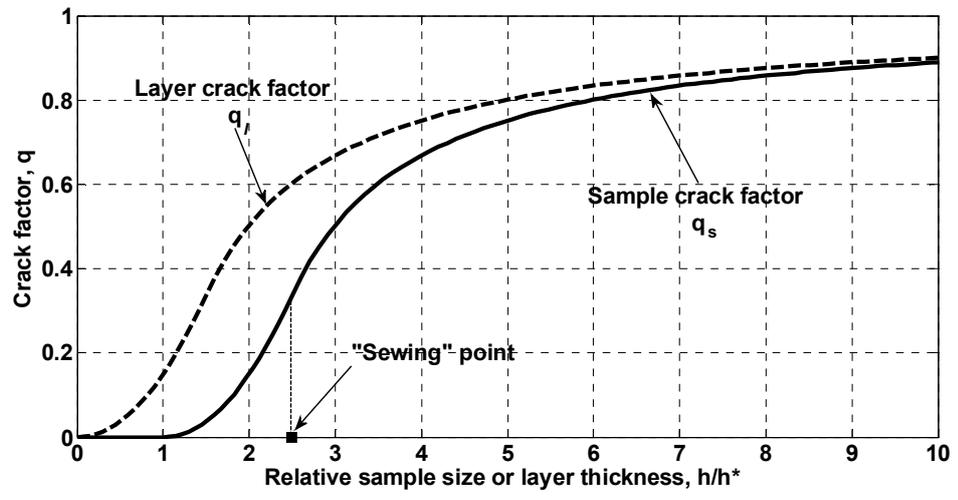

Fig.4

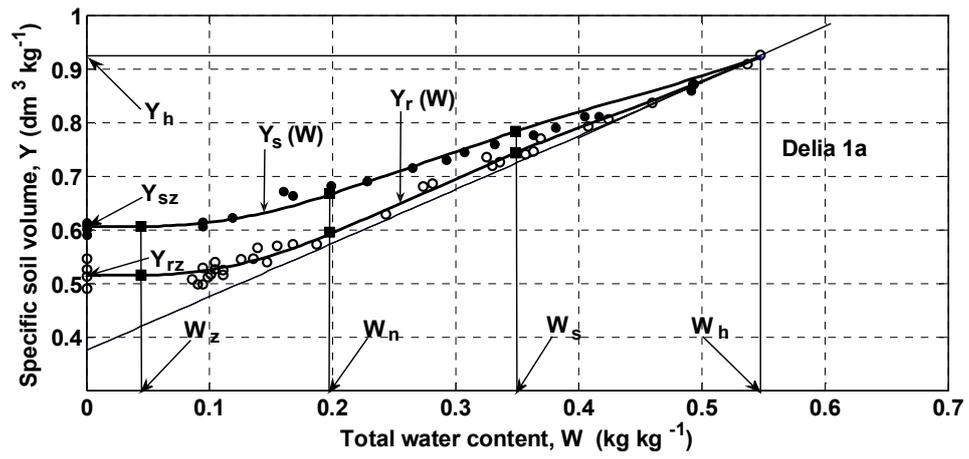

Fig.5

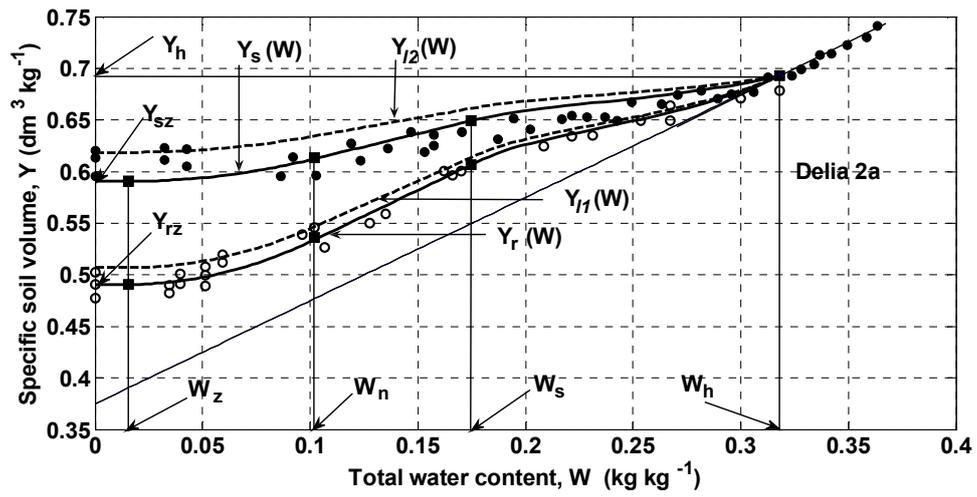

Fig.6

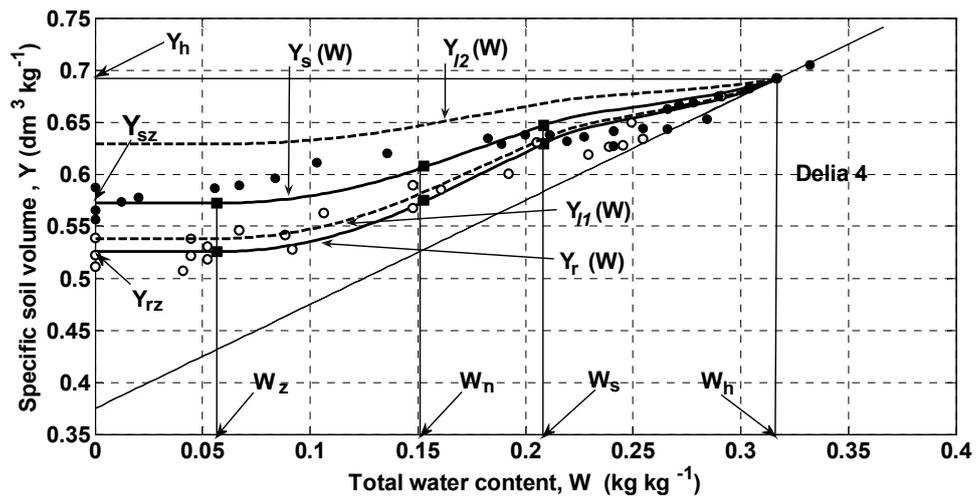

Fig.7

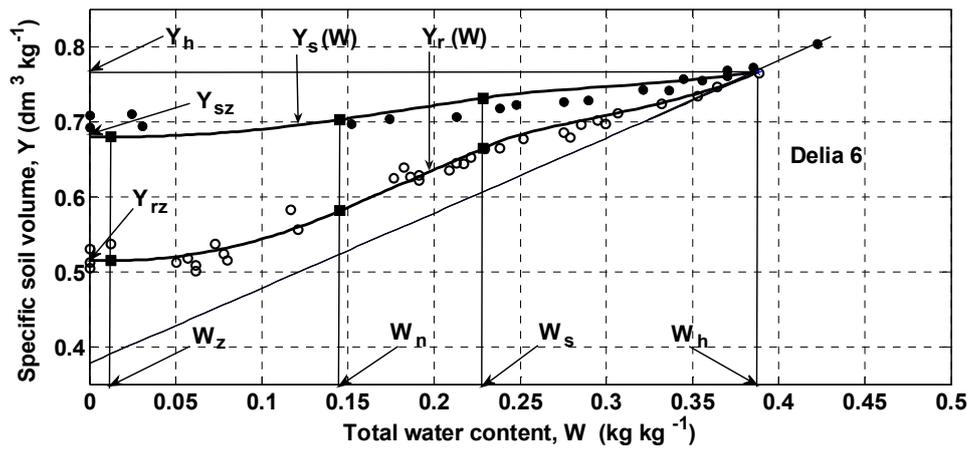

Fig.8

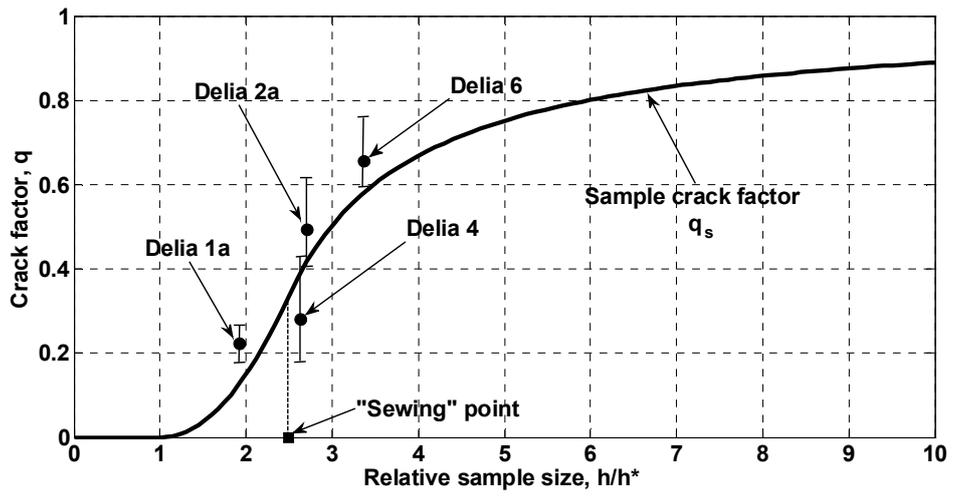

Fig.9

**Table 1.** Theoretically estimated general characteristics* of aggregate arrangement in soil volume for smallest and largest aggregate sizes in the typical ranges $0.02 < X_{min} < 0.07$ mm, and $2 < X_m < 5$ mm, respectively

| $X_{min}$ (mm) | $X_m$ (mm) | $l_{min}$ (mm) | $l_m$ (mm) | $h^*_o$ (cm) | $h^*$ (cm) | $l_{min}/X_{min}$ | $l_m/X_m$ |
|---|---|---|---|---|---|---|---|
| 0.02 | 2 | 1.03 | 146.63 | 1.2 | 5.3 | 50 | 73 |
| 0.02 | 5 | 3.42 | 685.32 | 4.8 | 5.3 | 171 | 137 |
| 0.07 | 2 | 1.47 | 114.29 | 1.3 | 4.8 | 21 | 57 |
| 0.07 | 5 | 5.05 | 542.40 | 5.2 | 4.6 | 72 | 108 |

* Minimum ($X_{min}$) and maximum ($X_m$) aggregate size at maximum swelling, mean distance between the smallest ($l_{min}$) and largest ($l_m$) aggregates at maximum swelling, rough approximation of the critical sample size ($h^*_o$), critical sample size ($h^*$), relative mean distance between the smallest ($l_{min}/X_{min}$) and largest ($l_m/X_m$) aggregates.

**Table 2.** Numerical primary experimental data and input parameters for reference shrinkage curve estimation

| Data source | Primary data[*] | | | Input parameters[**] | | | | |
|---|---|---|---|---|---|---|---|---|
| | $c$ | $s_1$ | $s_2$ | $Y_{rz}$ (dm$^3$kg$^{-1}$) | $W_h$ (kg kg$^{-1}$) | $\rho_s$ (kg dm$^{-3}$) | $P_z$ | $U_{lph}$ (dm$^3$kg$^{-1}$) |
| Crescimanno and Provenzano (1999, Table 1, Fig.4), soil Delia 1a | 0.57 | 0.35 | 0.08 | 0.5149 | 0.5476 | 2.6673 | 0 | 0 |
| As above, soil Delia 2a | 0.26 | 0.17 | 0.57 | 0.4907 | 0.3179 | 2.6673 | 0 | 0 |
| As above, soil Delia 4 | 0.22 | 0.19 | 0.59 | 0.5255 | 0.3166 | 2.6673 | 0 | 0 |
| As above, soil Delia 6 | 0.18 | 0.34 | 0.48 | 0.5150 | 0.3886 | 2.6428 | 0 | 0 |

[*] Clay content ($c$), silt content ($s_1$), sand content ($s_2$).
[**] Oven-dried specific volume ($Y_{rz}$), water content at maximum swelling ($W_h$), mean solid density ($\rho_s$), structural porosity in oven-dried state ($P_z$), specific lacunar pore volume at maximum swelling ($U_{lph}$).

**Table 3.** Estimated parameters of texture and structure* and intra-aggregate structure** of the four soils

| Soil | Possible $x_m$ range (mm) | Possible $X_{mz}$ range (mm) | Possible $x_n$ range (mm) | Possible $K$ range | $k$ | Possible $p$ range |
|---|---|---|---|---|---|---|
| Delia 1a | 0.072-0.085 | 4-5 | 0.014-0.015 | 1.034-1.038 | 0 | 0.130-0.642 |
| Delia 2a | 0.153-0.170 | 3.5-3.8 | 0.062-0.067 | 1.21-1.23 | 0 | 0.156-0.364 |
| Delia 4 | 0.150-0.170 | 4.7-5.7 | 0.064-0.070 | 1.15-1.18 | 0 | 0.128-0.370 |
| Delia 6 | 0.120-0.135 | 3.5-3.8 | 0.050-0.053 | 1.16-1.18 | 0 | 0.128-0.330 |

* Maximum (sand) grain size ($x_m$), maximum aggregate size in the oven-dried state ($X_{mz}$), mean soil-solids size ($x_n$).
** Aggregate/intra-aggregate mass ratio ($K$), lacunar factor ($k$), porosity of contributive silt and sand grains at *imagined* contact ($p$).

**Table 4.** Intermediate soil characteristics* participating in the $K$ and $k$ calculation

| Soil | $u_s$ | $u_S$ | $u_z$ | $u_h$ | $U_h$ (dm³kg⁻¹) | $U_i$ (dm³kg⁻¹) | $v_s$ | $v_z$ | $v_h$ | $f_{silt}$ | $p$ | $c_*$ |
|---|---|---|---|---|---|---|---|---|---|---|---|---|
| Delia 1a | 0.2550 | 0.1097 | 0.3402 | 0.6275 | 0.9225 | 0.0323 | 0.1633 | 0.2589 | 0.5816 | 0.8140 | 0.4165 | 0.3104 |
| Delia 2a | 0.3709 | 0.2745 | 0.4418 | 0.6855 | 0.6928 | 0.1242 | 0.1329 | 0.2306 | 0.5665 | 0.2297 | 0.2585 | 0.1673 |
| Delia 4 | 0.3719 | 0.2901 | 0.4933 | 0.6859 | 0.6915 | 0.1004 | 0.1152 | 0.2863 | 0.5576 | 0.2436 | 0.2145 | 0.0991 |
| Delia 6 | 0.3274 | 0.2685 | 0.4076 | 0.6637 | 0.7670 | 0.1140 | 0.0806 | 0.1901 | 0.5403 | 0.4146 | 0.1765 | 0.0833 |

* Relative volume of all solids of the intra-aggregate matrix ($u_s$), relative volume of non-clay solids of the intra-aggregate matrix ($u_S$), relative volume of the intra-aggregate matrix at shrinkage limit ($u_z$), relative volume of the intra-aggregate matrix at maximum swelling ($u_h$), specific volume of the intra-aggregate matrix at maximum swelling ($U_h$), contribution of the interface aggregate layer to the specific volume of soil aggregates ($U_i$), relative volume of the contributive-clay solids ($v_s$), relative volume of the contributive-clay matrix at shrinkage limit ($v_z$), relative volume of the contributive-clay matrix at maximum swelling ($v_h$), weight fraction of the silt grains among all silt and sand grains ($f_{silt}$), porosity of the contributive silt and sand grains at *imagined* contact ($p$), critical clay content ($c_*$).

**Table 5.** The estimated characteristics* of aggregate structure and arrangement in the four soils

| Soil | $x_m$ (mm) | $X_{mz}$ (mm) | $X_{min}$ (mm) | $X_m$ (mm) | $l_{min}$ (mm) | $l_m$ (mm) | $h^*_o$ (cm) | $h^*$ (cm) | $l_{min}/X_{min}$ | $l_m/X_m$ |
|---|---|---|---|---|---|---|---|---|---|---|
| Delia 1a | 0.075 | 4.50 | 0.015 | 5.51 | 3.47 | 808.26 | 5.3 | 6.0 | 231 | 146 |
| Delia 2a | 0.160 | 3.65 | 0.064 | 4.22 | 4.14 | 432.53 | 4.2 | 4.2 | 65 | 103 |
| Delia 4 | 0.160 | 5.00 | 0.067 | 5.59 | 5.92 | 670.98 | 6.3 | 4.4 | 88 | 120 |
| Delia 6 | 0.130 | 3.60 | 0.052 | 4.21 | 4.46 | 500.83 | 4.7 | 3.4 | 86 | 119 |

\* Maximum (sand) grain size ($x_m$), maximum aggregate size in the oven-dried state ($X_{mz}$), minimum aggregate size ($X_{min}=X_{minz}$), maximum aggregate size at maximum swelling ($X_m$), mean distance between the smallest ($l_{min}$) and largest ($l_m$) aggregates at maximum swelling, rough approximation of the critical sample size ($h^*_o$), critical sample size ($h^*$), relative mean distance between the smallest ($l_{min}/X_{min}$) and largest ($l_m/X_m$) aggregates.

**Table 6.** Some estimated shrinkage characteristics of large soil samples*

| Soil | $q_s$ | $\delta q_{s+}$ | $\delta q_{s-}$ | $S$ (dm$^3$kg$^{-1}$) | $h/h$* |
|---|---|---|---|---|---|
| Delia 1a | 0.223 | 0.042 | 0.042 | 0.777 | 1.926 |
| Delia 2a | 0.494 | 0.125 | 0.086 | 0.506 | 2.707 |
| Delia 4 | 0.280 | 0.150 | 0.103 | 0.720 | 2.637 |
| Delia 6 | 0.655 | 0.105 | 0.057 | 0.345 | 3.366 |

* Crack factor of sample ($q_s$), positive and negative standard errors of $q_s$ ($\delta q_{s+}$ and $\delta q_{s-}$), slope of sample shrinkage curve in the basic shrinkage range ($S$), relative initial sample size ($h/h$*).